\documentclass[pra, twocolumn, floatfix]{revtex4}
\usepackage{graphicx}
\usepackage{amsmath, amsfonts, amssymb, bm}
\usepackage{braket}
\usepackage{color}
\usepackage[utf8]{inputenc}
\usepackage{subfigure}

\begin{document}
\title{  
Photo fragmentation of large dimers into singly charged ions via the knockout mechanism }
\author{ B. Najjari$^1$ and A. B. Voitkiv$^2$ }
\affiliation{ $^1$ Institute of Modern Physics, Chinese Academy of Sciences, Lanzhou 730000, China \\  
$^2$ Institut for Theoretical Physics I, 
Heinrich Heine University of D\"usseldorf, 
\\ Universit\"atsstr. 1, 40225 D\"usseldorf, Germany }
\date{\today}
\begin{abstract} 

We study the fragmentation of very large dimers into 
two singly charged ions caused by absorption 
of a photon. In this process a photo electron emitted from one atom of the dimer has certain chances to hit the other atom knocking out one of its electrons. This results in the production of two singly charged ions and the consequent Coulomb explosion of the residual doubly charged system. 
We develop a theory of this process and apply it 
to calculate the fragmentation of 
$^4$He$_2$, $^7$Li - $^4$He and $^6$Li - $^4$He dimers.   
Our results for the helium dimer  
are in good agreement with available experimental data. 
For the Li-He dimers (where experimental data are still absent) our results predict that  
this fragmentation mechanism becomes much more efficient than in the case of He$_2$.  
We also show that, provided the recoil effects are of minor importance, 
the so called reflection approximation is very accurate also for large  dimers.  

\end{abstract}

\maketitle

\section{ Introduction } 

There has been theoretically predicted the existance of a number of very large-size dimers, like  
e.g. $^4$He$_2$, $^7$Li - $^4$He, $^6$Li - $^4$He, $^7$Li - $^4$He, Na - $^4$He, 
and Na - $^3$He (see \cite{dimers-theor} and references therein). 
They have tiny binding energies 
and very large space dimensions extending 
beyond several tens and even hundreeds of \AA. 

Two of these dimers, 
$^4$He$_2$ and $^7$Li - $^4$He, have been experimentally observed \cite{He2a}-\cite{He2b}, 
\cite{LiHe}. One of them,  
$^4$He$_2$, has already been extensively used in various experiments including those where the fragmentation of the $^4$He$_2$ dimer into 
He$^+$ ions,    
caused by the interaction with photons and charged particles, was explored 
\cite{He2_photon} - \cite{dimer-binding-exp}.  
A very large averaged distance between atoms in such a dimer makes its fragmentation to possess qualitatively different features compared to those known for "normal" diatomic molecules.  

In particular, the authors of \cite{He2_photon} reported on an experimental observation of an interesting process, in which the absorption of just a single photon 
with an energy of about $64$ eV by $^4$He$_2$ dimer leads 
to its fragmentation into He$^+$(1s) ions. 
They also proposed the mechanism governing this process: a photo electron emitted from one of the dimer atoms on its way out has certain chances to pass sufficiently close to the other atom knocking out one of its electrons. 
The residual He$^+$-He$^+$ system is unstable undergoing a Coulomb explosion. The energy threshold for this mechanism, which will be below referred to as {\it knockout}, is given by 
$\hbar \omega = 2 I_b $ ($\approx 49.2$ eV), 
where $ \hbar \omega $ is the photon energy and $I_b \approx 24.6$ eV is the ionization potential of He atom.    

Even though the results of \cite{He2_photon} 
were published more than a decade ago, we could not locate in the literature any theoretical treatment of this process 
\cite{prl-est-2010}. 
Besides, the knockout fragmentation mechanism 
should also be possible 
for other humungous dimers (like e.g. $^7$Li--$^4$He and $^6$Li--$^4$He) exposed to an electromagnetic radiation and 
an appropriate theory could be useful for future experimental activities in this field. 

Therefore, in the present article we make an 
attempt to develop a theoretical treatment of 
the fragmentation of large-size dimers into singly charged ions by absorption of a single photon via the knockout mechanism and apply it to describe the fragmentation of $^4$He$_2$, $^7$Li--$^4$He and $^6$Li--$^4$He dimers. 

\vspace{0.1cm} 

In order to put the present consideration into a broader perspective we note 
that the above mechanism is not the only one which can lead to the break up of a large dimer into two singly charged ions upon absorption of just a single photon.  

As was mentioned, in the case of the He$_2$ dimer the knockout mechanism becomes energetically allowed if the energy of the incident photon exceeds $\approx 49.2$ eV. 
However, with increasing photon energy other breakup 
mechanisms become possible as well.      

If the photon energy is sufficient 
for simultaneous ionization-excitation of He atom   
($ \hbar \omega > 65.4$ eV) 
the fragmentation can occur via 
de-excitation of the He$^+$ ion 
into the ground state  
in which the energy released is transferred to the neutral 
He atom that results in its ionization.  

When the excited He$^+$ is produced 
in a metastable state (e.g. 2s) the energy transfer occurs via Penning ionization. However, as has been known since 1960s \cite{icd-source}, 
the conversion of excitation of one atomic center 
into ionization of the other becomes much more efficient 
acquiring a significantly longer range when 
it is driven by the dipole-dipole interaction.  

For instance, a two-center dipole-dipole coupling  was found in \cite{i-a-a} to be the main driving mechanism of the inter-atomic Auger decay \cite{i-a-a-1}. 
Moreover, beginning with the work of \cite{icd-guys}, various relaxation mechanisms   
(generally termed intermolecular/interatomic Coulombic decay, ICD),  
driven mainly by a long-range 
dipole-dipole coupling, were shown to 
be of importance in many processes \cite{jahn}. 
In particular, it has been found in \cite{He2_photon_icd_exp} 
that simultaneous photo ionization-excitation of one of He  atoms with a consequent ICD dominates the total cross section for the 
breakup of He$_2$ into He$^+$ ions at photon energies 
$ \hbar \omega > 65.4$ eV 
(and at least up to $ \hbar \omega \lesssim 79$ eV).  

If the photon energy increases further and becomes 
sufficient to remove two electrons from He atom 
($ \hbar \omega > 79$ eV), 
yet another channel opens for the fragmentation 
of the He$_2$ dimer. 
Here, the absorption of a photon results in the production of 
He$^{2+}$ ion which then captures an electron from the helium atom. Since the capture may occur at rather small distances only, this mechanism is short ranged. To our knowledge, for photo fragmentation of He$_2$ this  mechanism has not yet been explored \cite{He2+}.     

In the case of Li-He dimers all the above photo 
fragmentation mechanisms are also possible. Moreover, due to the presence of two electronic shells in Li, new mechanisms arise. 
They begin with photo emission of one of the 
$K$-shell electrons of Li. The vacancy can then be filled either by the $2s$ electron of the residual Li$^+$ with the energy transferred to He via Penning ionization or by an electron from He with 
the energy excess absorbed by the 2s-electron of Li that results in its emission. The latter mechanism, which is very similar to the transfer ionization process in slow atomic collisions \cite{trans-ioniz}, 
is called electron transfer mediated decay (ETMD) \cite{ETMD}. 
The energy threshold for photo emission from the $K$-shell 
of Li is $ \hbar \omega \approx 64.4$ eV \cite{nist} which is also the effective threshold for these two fragmentation mechanisms \cite{1s2s2}.       

\vspace{0.25cm} 

Having briefly discussed  
the main photo fragmentation mechanisms for He$_2$ and Li-He dimers, resulting in the production of two singly charged ions, we now turn to the primary topic of the present paper: the knockout fragmentation mechanism.    

\vspace{0.1cm} 

The article is organized as follows. In Section II 
we develop two theoretical approaches to treat the fragmentation 
of a large-size dimer by absorption of a single photon via the knockout mechanism and discuss their inter-relation.   
In particular, in Section II we derive Eqs. 
(\ref{ga-11}) and (\ref{qa-34}) for the fragmentation cross sections which are used later for obtaining numerical results.    
In Section III we present and discuss these results 
for the fragmentation cross sections of $^4$He$_2$, $^6$Li - $^4$He and $^6$Li - $^4$He dimers.    
Section IV contains the main conclusions.    

\vspace{-0.35cm} 

\section{ Theoretical considerations }

In this section we shall present two different 
treatments for the fragmentation process, 
which both make use of the fact that the dimer under consideration has a very large space dimension.   

One of them is rather simple. It is 
based on the consideration of the dimer's geometry,  
operates from the onset with cross sections for elementary atomic processes and employs the so called reflection approximation, which relates the kinetic energy of the ionic fragments to 
an instantaneous size of the initial dimer. 
For brevity, we will refer to this approach as 
"geometric". 

The second treatment represents  
a more fundamental and rigorous approach which enables one to obtain a detailed description of the fragmentation process. 
It is based on the application of the quantum perturbation theory in the interaction of the "active" electrons of the dimer with the electromagnetic field and with each other. Within such an approach 
the non-zero contributions to the fragmentation process appear beginning with the second order terms in the perturbative expansion. We shall refer to this approach as "quantum".

\subsection{ "Geometric" approach } 

\subsubsection{ 
The production of two ions from two atoms by absorption of a single photon}

In our consideration we shall assume that both atoms of the dimer are initially at rest. Let $\frac{ d \sigma^{(A)}_{ph}}{ d \Omega_{\bm k} }$ 
be the cross section for photoionization of atom A 
of the dimer which is differential in the solid angle of the momentum ${\bm p}$ of the emitted electron;  
the absolute value of this momentum is given by 
$p^2/(2m_e) = \hbar  \omega + \varepsilon_i$, 
where $m_e$ is the electron mass, $ \hbar \omega$ is the photon energy and $\varepsilon_i $ is the electron energy in the ground state of atom A.  
Then the cross section $ \sigma^{2+}_{I} $ 
for the process, in which both atoms are singly ionized via photon absorption by atom A,  
can be estimated as 
\begin{eqnarray} 
\sigma^{2+}_{I}({\bm R}) = \int_{\Delta \Omega_{\bm p}} 
d \Omega_{\bm p} \, \, \frac{ d \sigma^{(A)}_{ph}}{ d \Omega_{\bm p} },    
\label{ga-1} 
\end{eqnarray}
where the integration runs over the element 
$\Delta \Omega_{\bm p} $ 
of the solid angle of the emitted electron.  
It depends on the internuclear vector $ {\bm R} $ 
of the dimer (assumed, for definiteness, to point towards atom B), $\Delta \Omega_{\bm p} = \Delta \Omega_{\bm p}({\bm R})$, and is supposed to have the property 
that the electron, which moves within it, ionizes atom B with the probablity equal to $1$ (otherwise the probability is zero). 
Since the distance between the atoms of the dimer is very large, $\Delta \Omega_{\bm p}$ has to be quite small and it is natural to estimate $\Delta \Omega_{\bm p}$ as 
\begin{eqnarray} 
\Delta \Omega_{\bm p} = \frac{ \sigma^{(B)}_{\text{e-2e}}(p)}{ R^2 }, 
\label{ga-2} 
\end{eqnarray} 
where $R = |{\bm R}|$ and $\sigma^{(B)}_{\text{e-2e}}(p)$ is the total cross section for ionization of atom B by the impact of an electron with incident momentum ${\bm p}$ 
($p = | \bm p|$). Since the solid angle 
$\Delta \Omega_{\bm p}$ is very small we can assume that within this angle the photo cross section 
$ \frac{ d \sigma^{(A)}_{ph}}{ d \Omega_{\bm p} }$ remains  essentially a constant,  
$ \frac{ d \sigma^{(A)}_{ph}}{ d \Omega_{\bm p} } \approx  
\frac{ d \sigma^{(A)}_{ph}}{ d \Omega_{ {\bm p}_{\bm R} } }$, 
where  ${\bm p}_{\bm R} = p \, \hat{\bm R} $ with 
$ \hat{\bm R} = \frac{ {\bm R} }{ R } $ 
(and thus one has $ \frac{ d \sigma^{(A)}_{ph}}{ d \Omega_{ {\bm p}_{\bm R} } } = \frac{ d \sigma^{(A)}_{ph}}{ d \Omega_{\bm R } }$).   
Then we obtain  
\begin{eqnarray} 
\sigma^{2+}_{I}({\bm R}) \approx \frac{ d \sigma^{(A)}_{ph}}{ d \Omega_{\bm R } } \, \, \, \frac{ \sigma^{(B)}_{\text{e-2e}}(p) }{ R^2 }. 
\label{ga-3} 
\end{eqnarray}

A similar consideration for the process, in which atom B is initially photoionized and the photo electron moving towards atom A knocks out one of its electrons, leads to the cross section 
\begin{eqnarray} 
\sigma^{2+}_{II}({\bm R}) \approx \frac{ d \sigma^{(B)}_{ph}}{ d \Omega_{{\bm R}'}  } \, \, \, 
\frac{ \sigma^{(A)}_{\text{e-2e}}(p')}{ R^2 }. 
\label{ga-4} 
\end{eqnarray}
Here, $ \frac{d \sigma^{(B)}_{ph}}
{ d \Omega_{ {\bm R}' } }$ is the differential cross section for photoinization of atom B with an emitted electron having an absolute value of the momentum $ p' $ 
($p'^2/2m_e = \hbar \omega + \epsilon_i$ where 
$\epsilon_i$ is the electron energy in the ground state of atom B) and moving along the vector 
$ {\bm R}' = - {\bm R}  $,  
$ {\bm p}'_{\bm R'} = - p' \, \, {\bm R}/R$. Further, 
$ \sigma^{(A)}_{\text{e-2e}}(p') $ 
is the total cross section for ionization of atom $A$ by the impact of an electron with incident momentum $ p' $.  

The total cross section for the production of two singly charged ions is given by the sum 
of (\ref{ga-3}) and (\ref{ga-4}): 
\begin{eqnarray} 
\sigma^{2+}({\bm R})  & = & \sigma^{2+}_{I}({\bm R}) + 
\sigma^{2+}_{II}({\bm R})  
\nonumber \\ 
& \approx & \! \!\! \frac{ 1 }{ R^2 }  
\bigg( 
\frac{ d \sigma^{(A)}_{ph}}{ d \Omega_{ \bm R } } \, \sigma^{(B)}_{\text{e-2e}}(p) \, + \,  
\frac{ d \sigma^{(B)}_{ph}}{ d \Omega_{ {\bm R}'} } \, \sigma^{(A)}_{\text{e-2e}}( p' ) \bigg). 
\label{ga-5} 
\end{eqnarray}

The above simply consideration can be extended to obtain 
the cross section for the production of two ions 
differential in the momenta 
${\bm p}_1$ and ${\bm p}_1$ 
of the emitted electrons:   
\begin{eqnarray} 
\frac{d \sigma^{2+}({\bm R})}{d^3 {\bm p}_1 \, d^3 {\bm p}_2} 
\!\!\!  & = & \!\!\! \frac{d \sigma^{2+}_{I}({\bm R})}{d^3 {\bm p}_1 \, d^3 {\bm p}_2} 
+ 
\frac{d \sigma^{2+}_{II}({\bm R})}{d^3 {\bm p}_1\,d^3 {\bm p}_2}   
\nonumber \\ 
& \approx & \!\!\!  \frac{ 1 }{ R^2 }  
\bigg(\!\!\! \frac{ d \sigma^{(B)}_{\text{e-2e}}({\bm p}) }{ d^3 {\bm p}_1 \, d^3 {\bm p}_2}  
\frac{ d \sigma^{(A)}_{ph}}{ d \Omega_{ \bm R } } \! + \!   
\frac{ d \sigma^{(A)}_{\text{e-2e}}({\bm p}') }{ d^3 {\bm p}_1 \, d^3 {\bm p}_2 } 
\frac{ d \sigma^{(B)}_{ph}}{ d \Omega_{ {\bm R}'} } \!\!\! \bigg).
\label{ga-6} 
\end{eqnarray}
Here, $ {\bm p} = p \, \hat{\bm R}$ and 
$ {\bm p}' = - p' \, \hat{\bm R}$ and the differential  
cross sections $\frac{ d \sigma^{(A)}_{\text{e-2e}}({\bm p}') }{ d^3 {\bm p}_1 \, d^3 {\bm p}_2 }$ and 
$\frac{ d \sigma^{(B)}_{\text{e-2e}}({\bm p}) }{ d^3 {\bm p}_1 \, d^3 {\bm p}_2}$ describe atomic ionization by the impact of an electron with momentum ${\bm p}'$ and ${\bm p}$, respectively. Note that both these cross sections are proportional to the delta-function 
$ \delta( p_1^2/2m_e + p_2^2/2m_e - \hbar \omega -  \varepsilon_i - \epsilon_i  ) $, 
which expresses the energy conservation in the 
"atomic" part of the fragmentation process.  

\subsubsection{ Cross sections for dimer fragmentation  }

Let us now consider the quantity  
\begin{eqnarray} 
\frac{ d \sigma_{fr}}{ d^3 {\bm R} } = \frac{d \sigma^{2+}({\bm R})}{d^3 {\bm p}_1 \, d^3 {\bm p}_2} \, \big| \psi_i({\bm R}) \big|^2,   
\label{ga-7} 
\end{eqnarray}
where $\psi_i({\bm R}) $ is the wave function describing 
the relative motion of atoms A and B (with masses $M_A$ and $M_B$, respectively) in the initial (ground) state of the A - B dimer.  
The system A$^+$ - B$^+$, into which the initial dimer 
is transformed, is unstable undergoing a Coulomb explosion.   
According to the reflection approximation the kinetic energy 
$E_{K}$ released by the ionic fragments due to the Coulomb explosion is related to the internuclear  distance $R$ 
at the time instance when the explosion begins: 
\begin{eqnarray} 
E_{K}  =  \frac{ Q_1 Q_2 }{ R },      
\label{ga-8} 
\end{eqnarray}  
where $Q_1 $ and $ Q_2$ are the ionic charges. 
If the time, which the photo electron needs to propagate from one center of the dimer to the other, is much shorter than the typical evolution times of the intermediate 
A$^+$-B and A-B$^+$ systems, the distance $R$ will 
practically coincide with the intantaneous size of the initial A-B dimer which the latter had when 
the photon was absorbed.  

Let us assume that the absolute value of 
the relative momentum 
$\bm K$ of the fragments, 
which result from the Coulomb explosion, 
is much larger than 
their recoil momenta acquired due to 
the electron emission. Under such a condition 
the momentum $\bm K$ will be directed along 
the internuclear vector $\bm R$ and, 
taking also into account 
(\ref{ga-8}), we get  
\begin{eqnarray} 
d^3 \bm R &=& 
\frac{ (Q_1 Q_2)^3 }{ \mu \, K \, E^4_{K} } 
\, \, d^3 {\bm K}.       
\label{ga-9} 
\end{eqnarray}   
Using (\ref{ga-6}) and (\ref{ga-7})-(\ref{ga-9}) we obtain 
the fragmentation cross section differential in the relative momentum of the ionic fragments and the momenta of the emitted electrons  
\begin{eqnarray} 
\frac{ d \sigma_{fr}}{ d^3 {\bm K} d^3 {\bm p}_1 d^3 {\bm p}_2 } 
\! = \! \frac{ Q_1 Q_2  }{ \mu \, K \, E^2_{K} } \, 
\bigg| \psi_i\bigg( \frac{ Q_1Q_2}{E_K} \hat{\bm K}  \bigg) \bigg|^2 
\nonumber \\ 
\times \bigg(\!\! \frac{ d \sigma^{(B)}_{\text{e-2e}}({\bm p}) }{ d^3 {\bm p}_1 d^3 {\bm p}_2}  \, 
\frac{ d \sigma^{(A)}_{ph}}{ d \Omega_{ \bm K } } +   
\frac{ d \sigma^{(A)}_{\text{e-2e}}({\bm p}')}{ d^3 {\bm p}_1 d^3 {\bm p}_2 } \,  
\frac{ d \sigma^{(B)}_{ph}}{ d \Omega_{ {\bm K}'} } \!\!\bigg),   
\label{ga-10} 
\end{eqnarray}
where now ${\bm p}  = p \hat{\bm K}$, 
${\bm p}'  = - p' \hat{\bm K}$ and 
the magnitudes of $p$ and $p'$ as well as the argument of the energy conserving delta-function 
have to be adjusted to account for 
the kinetic energy release $E_K$. In particular, 
the delta-function, which is contained in 
$\frac{ d \sigma^{(A)}_{\text{e-2e}}({\bm p}')}{ d^3 {\bm p}_1 d^3 {\bm p}_2 }$ and $\frac{ d \sigma^{(B)}_{\text{e-2e}}({\bm p}) }{ d^3 {\bm p}_1 d^3 {\bm p}_2}$,   
 becomes    
$ \delta\big( p_1^2/2m_e + p_2^2/2m_e + E_{K} - 
\varepsilon_i - \epsilon_i - \hbar \omega \big)$ that restores the energy balance in the process.  

By integrating the cross section (\ref{ga-10}) over the electron momenta we further get 
\begin{eqnarray} 
\frac{ d \sigma_{fr}}{ d^3 {\bm K}  } 
&=& \frac{ Q_1 Q_2  }{ \mu \, K \, E^2_{K} } \, 
\bigg| \psi_i\bigg( \frac{ Q_1Q_2}{E_K} \hat{\bm K}  \bigg) \bigg|^2 
\nonumber \\ 
&& \times \bigg( \!\! \sigma^{(B)}_{\text{e-2e}} \,
\frac{ d \sigma^{(A)}_{ph}}{ d \Omega_{ \bm K } } +   
\sigma^{(A)}_{\text{e-2e}} \, 
\frac{ d \sigma^{(B)}_{ph}}{ d \Omega_{ {\bm K}'} } \!\! \bigg)
\, ,   
\label{ga-11} 
\end{eqnarray}  
where $\sigma^{(A)}_{\text{e-2e}}$ and 
$\sigma^{(B)}_{\text{e-2e}}$ are the total cross section for electron impact ionization of atom A and B, respectively.  

Before we proceed to discuss the second approach 
a few more remarks may be appropriate. 

First, since the bound state of the A-B dimer is spherically symmetric the argument of the wave function 
$\psi_i $ actually does not depend on $\hat{\bm K}$. 

Second, in the energy balance we have neglected not only the binding energy of the initial dimer, which is tiny, 
but also the recoil energies of the ions. The latter can be done if they are much smaller than  
$ E_{K} $. If the recoil momenta do not noticeably exceed 
$1$ a.u., the recoil energies -- due to relatively very large masses of the nuclei -- will be in the range of meV:  
for instance, similar values of $ E_K $ would correspond to the beginning of the Coulomb explosion at the internuclear distances of the order of $10^4$ a.u. which is much larger than the size of any known  dimer. 

Third, if the kinetic energy release $E_K$ is not much smaller 
than the final electron energy $(p_1^2+p_2^2)/2m_e$ 
the cross sections $\sigma^{(A)}_{\text{e-2e}}$ and 
$\sigma^{(B)}_{\text{e-2e}}$ may be noticeably lower 
than their values for isolated atoms.   

\subsection{ Quantum approach }

\subsubsection{ The transition amplitude }

Keeping only the relevant terms, the second order transition amplitude reads 
\begin{eqnarray} 
a_{fi} &=& \bigg( -\frac{ i }{ \hbar } \bigg)^2 
\sum_{n} 
\int_{-\infty}^{+\infty} dt \, \big \langle \Psi_f(t) \big | \hat{ V }_{ee} \big | \Psi_n(t) \big \rangle 
\nonumber \\ 
&& \times \int_{-\infty}^{ t } dt' \, \big \langle \Psi_n(t') \big | \hat{ W }_{\gamma}(t) \big | \Psi_i(t') \big \rangle.  
\label{qa-1}  
\end{eqnarray}
In the above expression $\Psi_i$, $\Psi_n$ and $\Psi_f$ are 
the initial, intermediate and final states of the dimer, respectively, $\hat{ W }_{\gamma}(t) = \hat{ W }_{\gamma} \exp(- i \omega t) $ is the interaction of the "active" electrons of atom A and B with the electromagnetic field,   
$ \hat{ V }_{ee} $ the interaction between these  electrons,  
and the sum runs over the complete set of the intermediate states (including continuum ones).   
Using the Born-Oppenheimer approximation 
we present the states as  
\begin{eqnarray} 
\Psi_i(t) & = & \Phi_i(t) \, \,  
\, \psi_i \exp(- i E_i t/\hbar )
\nonumber \\ 
\Psi_n(t) & = & \Phi_n(t) \,\,\, \psi_n  
\exp(- i E_n t/\hbar ) 
\nonumber \\ 
\Psi_f(t) & = & \Phi_f(t) \,\,\, \psi_f 
\exp(- i E_f t/\hbar ) ,   
\label{qa-2}  
\end{eqnarray}
where $\Phi_i$, $\Phi_n$ and $\Phi_f$ refer to the initial, intermediate and final electron states 
whereas $\psi_i$, $\psi_n$ and $\psi_f$ denote  
the corresponding nuclear (molecular) states (which are determined by 
the respective electron configurations) and 
$E_i$, $E_n$ and $E_f$ are the molecular energies.  
We take the electron states according to 
\begin{eqnarray} 
\Phi_i(t) & = &  \phi_i \exp(- i \varepsilon_i t/\hbar ) 
\times \chi_i \exp( - i \epsilon_i t/\hbar ) 
\nonumber \\ 
\Phi_n(t) & = & \phi_{n_1} 
\exp(- i \varepsilon_{n_1} t/\hbar ) 
\times \chi_{n_2} \exp(- i \epsilon_{n_2} t/\hbar )
\nonumber \\ 
\Phi_f(t) & = & \phi_{f} \exp(- i \varepsilon_{f} t/\hbar ) 
\times \chi_f \exp(- i \varepsilon_{f} t/\hbar),   
\label{qa-3}  
\end{eqnarray}
where the symbols $\phi$ ($\chi$) refer to the states of 
the active electron of (or originated from) atom A (B) 
and $ \varepsilon_{i} $ ($ \epsilon_{i} $), 
$ \varepsilon_{n_1} $ ($ \epsilon_{n_2} $) 
and $ \varepsilon_{f} $ ($ \epsilon_{f} $) denote its initial, intermediate and final energies, respectively.   

Assuming that the interaction with the electromagnetic field $\hat{W}_\gamma$ 
is adiabatically switshed on and off at $t \to - \infty $ and $t \to +\infty $, respectively, we can perform in (\ref{qa-1}) the integrations over $t'$ and $t$. Taking also into account 
that $\hat{W}_\gamma$ is the sum of single-electron operators we obtain 
\begin{eqnarray} 
a_{fi} &=& 2 \pi i \, 
\delta(\varepsilon_f + \epsilon_f + 
E_f - \varepsilon_i - \epsilon_i - 
E_i - \hbar \omega) 
\nonumber \\ 
&& \times \big( a^I_{fi} + a^{II}_{fi} \big), 
\label{qa-4}  
\end{eqnarray}
where 
\begin{eqnarray} 
\!\!\!\!\!\! \! \!\! a^I_{fi} \! \! \! &=& \! \! \! \!\!   
\sum_{n, n_1} \!\!
\frac{ \big \langle \psi_f \phi_f \chi_f \big | \hat{ V }_{ee} \big | \psi_n \phi_{n_1} \chi_{i} \big \rangle 
\big \langle \psi_n \, \phi_{n_1}  \big | 
\hat{ W }^A_{\gamma} \big | \phi_{i} \, \psi_i \big \rangle }{ \varepsilon_{n_1} - \varepsilon_{i} - \hbar \omega + 
E_n - E_i - i 0 } 
\label{qa-5}  
\end{eqnarray}
and 
\begin{eqnarray} 
\!\!\!\!\!\! \! \!\! a^{II}_{fi} \!\!\! & = & \!\!\! \!\! 
\sum_{n, n_2} \!\!
 \frac{ \big \langle \psi_f \phi_f \chi_f \big | \hat{ V }_{ee} \big | \psi_n \phi_{i} \chi_{n_2} \big \rangle 
\big \langle \psi_n \, \chi_{n_2}  \big | 
\hat{ W }^B_{\gamma} \big | \chi_{i} \, \psi_i  \big \rangle }{ \epsilon_{n_2} - \epsilon_{i} - \hbar \omega + 
E_n - E_i - i 0 } .  
\label{qa-6}  
\end{eqnarray}
In the above expressions the sum over $n_1$ and $n_2$ implies also the integration over the electron continuum states whereas the sum over $n$ runs over all molecular states $\psi_n$ 
(which themselves depend on the corresponding electron configurations). 

Let us first consider the amplitude $a^I_{fi}$ describing the process in which the dimer fragmentation is caused by photo absorption on atom A. Assuming that the photon energy is well above the electron binding energy in atom A we neglect the contribution from all intermediate bound electron states obtaining 
\begin{eqnarray} 
a^I_{fi} \!\!\!&=& \!\!\! \!
\sum_{n} \!\! \int \!\!\! \frac{V d^3 {\bm p} }{ (2 \pi \hbar)^3 }  
\nonumber \\ 
&& \!\!\!\frac{ \big \langle \psi_f \phi_f \chi_f \big | \hat{ V }_{ee} \big | \psi_n \phi_{\bm p} \chi_{i} \big \rangle 
\big \langle \psi_n \phi_{\bm p} \big | 
\hat{ W }^A_{\gamma} \big | \phi_{i} \psi_i \big \rangle }{ \varepsilon_{\bm p} - \varepsilon_{i} - \hbar \omega + 
E_n - E_i - i 0 },  
\label{qa-7}  
\end{eqnarray} 
where $V$ is the normalization volume for the electron emitted  from atom A, ${\bm p}$ and  
$ \varepsilon_{\bm p} = p^2/2m_e$ 
are its momentum and energy, respectively, 
and $\psi_n$ are the states of the 
corresponding molecular A$^+$-B ion.      

The quantity 
$\big \langle \phi_f \chi_f \big | \hat{ V }_{ee} \big |  \phi_{\bm p} \chi_{i} \big \rangle = 
\big \langle \phi_{{\bm p}_1} \chi_{{\bm p}_2} \big | \hat{ V }_{ee} \big |  \phi_{\bm p} \chi_{i} \big \rangle $, 
where $\hat{ V }_{ee} $ 
is the interaction between the electrons, 
describes ionization of atom B by the impact of an electron with momentum ${\bm p} $, and $\phi_{{\bm p}_1}$, and $\chi_{{\bm p}_2}$ are the final electron states with 
momenta ${\bm p}_1$ and ${\bm p}_2$. The states 
$ \chi_i $, $\phi_{{\bm p}_1}$, and $\chi_{{\bm p}_2}$ depend on the coordinates of the electrons with respect to the nucleus of atom B 
whereas the state $\phi_{{\bm p}}$ depends on the coordinates of the electron with respect to the nucleus of atom A. Taking into account 
that at large internuclear distances $R$ the state $ \phi_{{\bm p}} \equiv  
\phi_{{\bm p}}({\bm r} - {\bm R}_A) = 
\phi_{{\bm p}}({\bm r} - {\bm R}_B + {\bm R}_B - {\bm R}_A) = \phi_{{\bm p}}({\bm r} - {\bm R}_B + {\bm R})$ can be approximated as $\exp(i {\bm p} \cdot {\bm R}/\hbar) \,  
\phi_{{\bm p}}({\bm r} - {\bm R}_B)$ we obtain 
\begin{eqnarray} 
\!\!\!\!\!\!\!\!\!\!\!\big \langle \phi_{{\bm p}_1} \! \chi_{{\bm p}_2} \big | \hat{ V }_{ee} \big |  \phi_{\bm p} \chi_{i} \big \rangle 
\!\!\!&=& \!\!\! \exp(i {\bm p} \! \cdot \! {\bm R}/\hbar )     
\big \langle \phi_{{\bm p}_1} \!\chi_{{\bm p}_2} \!\big | \hat{ V }_{ee}  \big |  
\phi_{\bm p} \chi_{i} \big \rangle,   
\label{qa-9} 
\end{eqnarray}
where on the right-hand side of Eq. (\ref{qa-9}) 
the electron coordinates in the all states 
are with respect to the nucleus of atom B.  

Noting that the state $ \phi_{\bm p}({\bm x})$ 
can be written as 
$ \phi_{\bm p} \equiv \langle {\bm x} | {\bm p} \rangle $ 
let us consider the operator 
\begin{eqnarray}
\hat O = \int \!   
d \Omega_{\bm p} \, 
\exp\big(i {\bm p} \cdot {\bm R }/\hbar \big) \, \,  
\big|{\bm p} \big \rangle \big \langle  
{\bm p} \big | ,  
\label{qa-11} 
\end{eqnarray}
which enters the integrand in Eq. (\ref{qa-7}). 
By expanding in (\ref{qa-11}) the plane wave 
into partial waves we obtain 
\begin{eqnarray}
\hat O = 4 \pi \sum_{\lambda \mu} i^\lambda 
j_\lambda (p R/\hbar) Y_{\lambda \mu}(\hat {\bm R}) \int \!d \hat{\bm p} \; Y_{\lambda\mu}^* (\hat{\bm p}) \, 
\big| {\bm p} \big \rangle \big \langle {\bm p} \big|,   
\label{qa-12} 
\end{eqnarray}
where $j_\lambda$ are the spherical Bessel functions. Representing the vector $\big| \bm p \big \rangle$ as  
\begin{eqnarray} 
\big|{\bm p} \big> = \sum_{lm} 
\big| p \, l m \big \rangle {Y}_{lm}(\hat{\bm p})
\label{qa-13}
\end{eqnarray} 
the projection operator 
$ \big| {\bm p} \big \rangle \big \langle {\bm p} \big|$ becomes 
\begin{eqnarray} 
\big| {\bm p} \big \rangle \big \langle {\bm p} \big| 
= \sum_{lm \, l'm'} 
\big| p \, l m \big \rangle \big \langle p \, l' m' \big| 
\, {Y}^*_{l'm'}(\hat{\bm p}) {Y}_{lm}(\hat{\bm p})   
\label{qa-14} 
\end{eqnarray}
Using the relation \cite{v-m-h}  
\begin{eqnarray}
Y^*_{l_{1} m_{1}}(\hat {\bm a}) Y_{l_{2} m_{2}}(\hat{\bm a}) 
&=& \sum_{L M} {\cal A}_{l_1 m_1 : l_2 m_2}^{L M}  \;Y_{L M}(\bm {\hat a} ), 
\label{qa-15}   
\end{eqnarray}
where 
\begin{eqnarray}
{\cal A}_{l_1 m_1 : l_2 m_2}^{L M}
\!\! &=& \!\!\! (-1)^{m_1}  \,   
[l_1\;l_2\;L] \,  
C_{l_{1} 0\; l_{2} 0}^{L0} \; 
C_{l_1 -m_1 \; l_2 m_2}^{L M},  
\nonumber    
\end{eqnarray}
$[l_1\;l_2\;L] = \sqrt{\frac{\left(2 l_{1}+1\right)\left(2 l_{2}+1\right)}{4 \pi(2 L+1)}}$ 
and $C_{a_1 b_1 \; a_2 b_2}^{A B}$ are the Clebsch-Gordan coefficients, we obtain 
\begin{eqnarray}
\big|\bm p \big\rangle \big \langle \bm p \big| 
\! \! \! &=& \! \! \! \sum_{lm\;l'm'} \big| p \, lm \big \rangle \big \langle p \, l'm' \big| 
\sum_{L M} \! {\cal A}_{l m : l' m'}^{L M}  Y_{L M}(\hat {\bm p} )
\label{qa-16}
\end{eqnarray}
and thus the operator $\hat O$ reads   
\begin{eqnarray}
\hat O \!\! & = & \!\! 4 \pi \!\!\! \sum_{lm\;l'm'}\!\!\! 
\big| p \, lm \big \rangle \big \langle p \, l' m'\big| 
\! 
\nonumber \\ 
&& \times \sum_{L M} \! {\cal A}_{l m : l' m'}^{L M} \, \! 
i^L \! j_L(p R/\hbar) Y_{LM}(\hat{\bm R}).   
\label{qa-17} 
\end{eqnarray} 
Taking into account that for any function $f(p)$, 
which does not possess poles close to the real axis and behaves "reasonably" at infinity, one has 
(see e.g. \cite{b-s})
\begin{eqnarray} 
\lim_{ p_0 R \gg \hbar } \int_{0}^{\infty} \!\!d{p}\; p^2 i^L j_L(pR/\hbar) {f(p) \over p^2 - p_0^2 -i\eta }  
\nonumber \\ 
= {\pi \hbar \over 2} f(p_0)  
\frac{ \exp(i p_0 R/\hbar) }{ R }, 
\label{qa-18}
\end{eqnarray}
we obtain 
\begin{eqnarray} 
&& \int d^3 {\bm p} \, 
\exp(i {\bm p} \cdot {\bm R}/\hbar ) \, 
\frac{ \big|{\bm p} \big \rangle \big \langle {\bm p} \big | }
{ \varepsilon_{\bm p} - \varepsilon_{i} - \hbar \omega + 
E_n - E_i - i 0 } 
\nonumber \\ 
&& = 4 \pi^2 \hbar \, m_e \, 
\frac{ \exp(i \, p^{n}_0 \, R/\hbar ) }{ R } \, 
\big | { \bm p}^n_{0 \bm R}  \big \rangle  
\big \langle {\bm p}^n_{0 \bm R} \big |, 
\label{qa-19}  
\end{eqnarray}
where $ {\bm p}^n_0 = p^n_0 \hat{\bm R } $ with 
$p^{n}_0 = \sqrt{ 2m_e \, (\varepsilon_i + 
\hbar \omega + E_i - E_n) } $  
is the momentum of the electron emitted from atom A 
and $|{\bm p}^n_{0 \bm R} \rangle $ denotes its state.  

The absolute value $p^{n}_0$ of the electron momentum depends on the  
state of the transitory A$^+$-B dimer. 
The electronic, vibrational and rotational levels 
in a molecule scale in general as  
$ \sim 1 $, $ \sim \sqrt{m_e/ \mu} $ and 
$ \sim m_e/\mu $, respectively, where $\mu $ is the reduced nuclear mass \cite{LL}. Therefore, we can approximate 
$|{\bm p}^n_{0 \bm R} \rangle \approx |{\bm p}_{0 \bm R} \rangle$, 
where $ {\bm p}_0 = p_0 \hat{\bm R } $ with 
$p_0 = \sqrt{ 2m (\varepsilon_i + \omega) } $. 
Besides, one has 
$(p^{n}_0  - p^{m}_0) \, R/\hbar  \approx 
\frac{(E_m - E_n) }{\hbar }
\frac{ m_e R }{ p_0 } $ that can be interpreted as the ratio between the time 
$ T = m_e R/p_0 = R/v_0$, which the electron needs  for propagating from A to B, and the typical molecular evolution times  
$\tau \simeq \hbar/(E_m - E_n)$ in the transient A$^+$-B dimer. 

Provided $T \ll \tau$, the term  
$\exp(i \, p^{n}_0 R/\hbar)$ in (\ref{qa-20}) can be replaced by 
$\exp(i \, p_0 R/\hbar)$ and the summation over the states of the transient dimer is easily performed:  
\begin{eqnarray} 
&& \sum_n \big| \psi_n \big \rangle 
\big \langle \psi_n \big| \, 
\exp(i \, p^{n}_0 R/\hbar) 
\approx \exp(i \, p_0 R/\hbar).     
\label{qa-20}
\end{eqnarray}
Using Eqs. (\ref{qa-19})--(\ref{qa-20}) we obtain 
for the transition amplitude 
\begin{eqnarray} 
a^I_{fi} &=& 
\frac{ 4 \pi^2  V m_e \hbar }{ (2 \pi \hbar)^3 } 
\int \!\!d^3 {\bm R} \,\,  
\psi^*_f({\bm R}) \,  
J({\bm R}) \, \psi_i({\bm R}),  
\label{qa-21}  
\end{eqnarray}   
where 
\begin{eqnarray} 
J({\bm R}) &=&   
\frac{ \exp(i {\bm p}_{0} \cdot {\bm R }/\hbar) }{ R } 
\nonumber \\ 
&& \!\!\! \times \big \langle {\bm p}_1, {\bm p}_2 \big | 
\hat{V}_{ee} \big | { \bm p}_{0 \bm R}, \chi_i \big \rangle  
\big \langle {\bm p}_{0 \bm R} \big | 
\hat{ W }^A_{\gamma} \big | \phi_i \big \rangle.      
\label{qa-22}  
\end{eqnarray}
Note that, since $ {\bm p}_{0} \cdot {\bm R } = 
{\bm p}_{0} \cdot {\bm R }_B + (-{\bm p}_{0}) \cdot {\bm R }_A $, the "retardation" term 
$\exp(i {\bm p}_{0} \cdot {\bm R }/\hbar)$ in (\ref{qa-22}) can be viewed as describing the recoil of the nuclei of A and B caused by photo electron emission from A and the electron impact on B, respectively. However, this obviously does not fully take 
into account the momentum balance between the electrons 
and the nuclei in the process under consideration. 
Since on the molecular time scale the scattered and emitted electrons leave the molecule essentially instantaneously, 
the corresponding recoil effect on the nuclear motion can be taken within the sudden approximation by multiplying 
the molecular wave function $\psi_i$ by the factor 
$\exp( - i({\bm p}_1 + {\bm p}_2) \cdot {\bm R}_B )$. Then the quantity $J({\bm R})$ is substituted by   
\begin{eqnarray} 
\tilde{J}({\bm R}) &=& 
\frac{ \exp(i {\bm P}_{r} \cdot {\bm R }/\hbar) }{ R }  
\nonumber \\ 
&& \!\!\! \times \big \langle {\bm p}_1, {\bm p}_2 \big | 
\hat{V}_{ee} \big | { \bm p}_{0 \bm R}, \chi_i \big \rangle  
\big \langle {\bm p}_{0 \bm R} \big | 
\hat{ W }^A_{\gamma} \big | \phi_i \big \rangle,     
\label{qa-23}  
\end{eqnarray}
where the recoil momentum 
${\bm P}_{r} = {\bm p}_{0} - 
\frac{ M_A }{M_A + M_B} ({\bm p}_1+{\bm p}_2)$ 
affects the relative motion of the nuclei 
of the dimer.   
(Besides, the center-of-mass motion 
of the A$^+$-B$^+$ system  
asquires the recoil momentum 
${\bm P}_{cm} = -({\bm p}_1+{\bm p}_2)$.)    
Thus, the amplitude $a^I_{fi}$ becomes  
\begin{eqnarray} 
a^I_{fi} \!\!  
&=& \!\! \frac{ 4 \pi^2  V m_e \hbar }{ (2 \pi \hbar)^3 } 
\int \!\!d^3 {\bm R} \, \,    
\psi^*_f({\bm R})  
\, \tilde{J}({\bm R}) \,\, \psi_i({\bm R}).    
\label{qa-24}  
\end{eqnarray} 
 
\vspace{0.25cm} 

A similar consideration for the channel, in which the dimer fragmentation occurs due to photoabsorption on atom B, yields the amplitude 
\begin{eqnarray} 
a^{II}_{fi} \!\!  
&=& \!\! \frac{ 4 \pi^2  V m_e \hbar }{ (2 \pi \hbar)^3 } \, 
\int \!\!d^3 {\bm R} \, \, 
\psi^*_f({\bm R}) 
\, \tilde{J}'({\bm R}) \, \psi_i({\bm R}),   
\label{qa-26}  
\end{eqnarray} 
where 
\begin{eqnarray} 
\tilde{J}'({\bm R}) & = & \frac{ \exp(i \, \, {\bm P}'_{r} \cdot {\bm R }/\hbar) }{ R } 
\nonumber \\   
&& \! \! \! \times \big \langle {\bm p}_1, {\bm p}_2 \big | 
\hat{V}_{ee} \big | { \bm p}'_{0 \bm R}, \phi_i \big \rangle \, \big \langle {\bm p}'_{0 \bm R} \big | 
\hat{ W }^B_{\gamma} \big | \chi_i \big \rangle.  
\label{qa-26a} 
\end{eqnarray}
Here, $\hat{ W }^B_{\gamma} $ is the interaction of the photon field with atom B, $ \big | { \bm p}'_{0 \bm R} \big \rangle$ is the state of the photoelectron emitted from atom $B$ with momentum 
${\bm p}'_{0} = - \sqrt{2m_e \, (\epsilon_i + \hbar \omega)} 
\, \, \hat{\bm R} = - p'_0 \, \hat{\bm R} $ and  
${\bm P}'_{r} = {\bm p}'_{0} + 
\frac{ M_B }{M_A + M_B} ({\bm p}_1+{\bm p}_2)$. 
(Besides, as before, the center-of-mass motion of the A$^+$- B$^+$ system  
asquires the recoil momentum 
${\bm P}_{cm} = -({\bm p}_1+{\bm p}_2)$.)  

The total transition amplitude 
for the fragmentation is then given by 
\begin{eqnarray} 
a_{fi} \!\! &=& \! \! - 2 \pi  \,  
\delta\bigg( \!\! \frac{p_1^2}{2 m_e} + \frac{p_2^2}{2 m_e} + E_f - \varepsilon_i - \epsilon_i - E_i - \hbar \omega 
\!\! \bigg) 
\nonumber \\ 
&& \!\!\! \times \big( a^I_{fi} + a^{II}_{fi} \big), 
\label{qa-27}  
\end{eqnarray}
where $E_i$ is the energy of 
the ground state of the A-B dimer, $E_f $ 
is the energy of the final A$^+$ - B$^+$ system. 

\subsubsection{ The fragmentation cross sections }

The fully differential cross section for 
the fragmentation is given by 
\begin{eqnarray} 
\frac{ d \sigma_{fr} }{ d^3 {\bm K} \, 
d^3 {\bm p}_1 \, d^3 {\bm p}_2 } & = & 
\lim_{T \to \infty}  \, \, \frac{ 1 }{ j_{ph} } \frac{ |a_{fi}|^2 }{ T } 
\frac{ V_r  V_{e1} V_{e2} }{ (2 \pi \hbar)^{9} } 
\nonumber \\ &=& 
\frac{ 1 }{ j_{ph} }  
\frac{ V_r  V_{e1} V_{e2} }{ (2 \pi \hbar)^{9} }  
\, \big( 2 \pi \big)^2  
\frac{ 1 }{ 2 \pi \hbar } \, \, \, 
\big| a^I_{fi} + a^{II}_{fi} \big|^2 \, \, 
\nonumber \\ 
&  & \times 
\delta\big( {\cal E}_f- {\cal E}_i \big).  
\label{qa-28}  
\end{eqnarray}
Here, $j_{ph}$ is the photon flux, $V_{e1}$ and $V_{e2}$ are the normalization volumes for the emitted electrons, and $V_r$ is the normalization volume for the final state $\psi_f$ of the relative motion of the A$^+$-B$^+$ system. 
This state is a continuum state,  
$\psi_f = \psi_{\bm K}({\bm R})$, describing 
the relative motion with the momentum ${\bm K}$ of 
the A$^+$ and B$^+$ fragments. 
Further, ${\cal E}_i = \varepsilon_i + \epsilon_i + E_i +  \hbar \omega $ and 
$ {\cal E}_f = \frac{p_1^2}{2 m_e} + 
\frac{p_2^2}{2 m_e} + \frac{{\bm K}^2}{2 \mu }$ 
are the initial and final total energies, 
respectively \cite{en-cons}.  

The cross section (\ref{qa-28}) with the amplitudes 
$a^I_{fi} $ and $ a^{II}_{fi}$, given by Eqs. 
(\ref{qa-24})-(\ref{qa-26}), contains 
a detailed information about the fragmentation process. 
However, since it is quite difficult to evaluate, 
significant simplifications are very desirable. 

\subsubsection{ Simplified fragmentation cross sections }

The amplitudes (\ref{qa-24}) and (\ref{qa-26}) were derived assuming that the size of the dimer is large enough 
($ p_0 R \gg \hbar $, $ p'_0 R \gg \hbar $).  
This point can be exploited further by recalling that, 
because of this large size, 
electron impact ionization of atoms A and B 
is caused by photo electrons which move exactly in the opposite directions. Therefore, one can expect that the angular pattern of the emitted electrons in these two reaction pathways will have little overlap. This means that, provided we are not interested in subtle details of the electron and ion distributions, we can approximate 
\begin{eqnarray} 
| a^I_{fi} + a^{II}_{fi} |^2 \approx 
| a^I_{fi} |^2 + | a^{II}_{fi} |^2.  
\label{qa-29} 
\end{eqnarray}  

Another significant simplification becomes possible 
if the magnitudes of the recoil momenta ${\bm P}_r$ and ${\bm P}'_r$ of the heavy fragments (which the latter acquire 
due to the emission and scattering of the electrons)  
are significantly 
smaller than the absolute value of the  
momentum $ {\bm K} $ due to the Coulomb explosion of the 
residual A$^+$-B$^+$ system. In such a case 
the influence of the recoil momenta on the motion of the heavy fragments (which can be termed as recoil effects) 
is expected to be of minor importance. 
Moreover, since the energies, which the heavy fragments 
gain due to the recoil effects, scale      
as $ \sim {\bm P}^2_r$ ($ \sim {{\bm P}'}^2_r$)
whereas their energy due to the Coulomb repulsion 
is $ \sim {\bm K}^2$, it is seen that 
-- provided $|{\bm P}_r| \ll |{\bm K}| $ and 
$| {\bm P}'_r| \ll |{\bm K}| $ -- the recoil effects  
would affect the kinetic 
energy release spectrum (even) much less than the angular distributions 
of the fragments.  
  
For instance, even if the kinetic energy release 
$E_K = K^2/2 \mu $ in the fragmentation of $^4$He$_2$, 
$^6$Li - $^4$He or $^7$Li - $^4$He dimers 
is as small as $\approx 0.1$ eV, 
the absolute value of ${\bm K}$, $ K = |{\bm K}|$, 
still remains rather large, $ K \approx 5 - 6$ a.u. 
Therefore, if the absolute values of 
the recoil momenta ${\bm P}_r$ and ${\bm P}'_r$ 
do not noticeably exceed $1$ a.u. \cite{competition}, 
their effect on the energy spectra are expected 
to be quite weak up to very low energies 
which correspond to 
large internuclear distances (for instance, 
$ E_K \gtrsim 0.1$ eV corresponds to 
$R \lesssim 150 $ \AA).   
 
Taking all this into account, when calculating the energy distributions, $\exp(i \, \, {\bm P}_{r} \cdot {\bm R }/\hbar)$ and $\exp(i \, \, {\bm P}'_{r} \cdot {\bm R }/\hbar)$  
in (\ref{qa-23}) and (\ref{qa-26a})  
can be approximated by $1$. 

\vspace{0.15cm} 

Further, in the radial Schr\"odinger equation for the relative motion of the A$^+$-B$^+$ system the "centrifugal" part, 
$ \frac{ \hbar^2 l(l+1) }{ 2 \mu R^2} $,  
of the effective interaction  
$V_l(R) = \frac{ Q_1 Q_2 } {R} + \frac{ \hbar^2 l(l+1) }{ 2 \mu R^2} $ is not important when  
the classical turning distance $R_c =\frac{ Q_1 Q_2 }{E_K }  $ is much larger than the distance  
$R_l = \frac{ \hbar^2 l(l+1)}{ 2 \mu Q_1 Q_2} $ at which the quantity $\frac{ \hbar^2 l(l+1) }{ 2 \mu R^2} $ 
becomes equal to the Coulomb potential energy $\frac{ Q_1 Q_2 } {R}$.   
For the ratio $ \frac{ R_l }{ R_c } $ we obtain 
$ \frac{ R_l }{ R_c } = \frac{ \hbar^2 l(l+1) E_K}{ 2 \mu Q_1^2 Q_2^2} = 
\frac{l(l+1)}{4 \eta^2} $, where  
$\eta = \frac{ Q_1 Q_2 }{ \hbar } \sqrt{\frac{ \mu }{ 2 E_K} } = \frac{ Q_1 Q_2 }{ \hbar v } $ is the Sommerfeld parameter 
($v$ is the relative velocity of the heavy fragments). 
In our case this parameter is very large: for instance, 
for kinetic energy release 
$E_K$ varying in the range $0.1$--$5$ eV 
one has $ \eta \simeq 100$ - $700$ for the $^4$He$^+$-$^4$He$^+$ system (and even somewhat larger for 
$^6$Li$^+$-$^4$He$^+$ and $^7$Li$^+$-$^4$He$^+$). 
This not only means that 
the relative motion of the heavy fragments A$^+$ and B$^+$ is "almost" classical but also that the radial wave functions  describing this motion will essentially be independent 
of $l$ up to $l$ as large as $l \simeq 70$.  

\vspace{0.15cm} 

Performing the expansion of the terms, which enter   
the amplitudes $a^{I}_{fi}$ and $a^{II}_{fi}$, 
into partial waves and taking into account what 
has been said in the previous two paragraphs, 
one can show that these amplitudes can be factorized into three parts describing three steps of the fragmentation process: 
photo absorption, electron impact ionization and the molecular transition,    
\begin{eqnarray} 
a^I_{fi} \!\!  
&=& \!\! \frac{ 4 \pi^2  V m_e \hbar }{ (2 \pi \hbar)^3 } \, 
\bigg [\! \int \!\!d^3 {\bm R} \, \, \psi^*_f({\bm R})  
\, \frac{ 1 }{ R } \,\,  \psi_i({\bm R}) \bigg] \,  
\nonumber \\ 
&& \!\!\! \times \big \langle {\bm p}_1, {\bm p}_2 \big | 
\hat{V}_{ee} \big | { \bm p}_{0 \bm K}, \chi_i \big \rangle \times \big \langle {\bm p}_{0 \bm K} \big | 
\hat{ W }^A_{\gamma} \big | \phi_i \big \rangle   
\label{qa-30}  
\end{eqnarray}  
and 
\begin{eqnarray} 
a^{II}_{fi} \!\!   
&=& \!\! \frac{ 4 \pi^2  V m_e \hbar }{ (2 \pi \hbar)^3 } \, 
\bigg [\! \int \!\!d^3 {\bm R} \, \, 
\psi^*_f({\bm R})  
\, \frac{ 1 }{ R } \,\, \psi_i({\bm R}) \bigg] \,  
\nonumber \\ 
&& \!\!\! \times \big \langle {\bm p}_1, {\bm p}_2 \big | 
\hat{V}_{ee} \big | { \bm p}'_{0 \bm K}, \phi_i \big \rangle \times \big \langle {\bm p}'_{0 \bm K} \big | 
\hat{ W }^B_{\gamma} \big | \chi_i \big \rangle,    
\label{qa-31}  
\end{eqnarray}
where 
${ \bm p}_{0 \bm K} = p_0 \hat{\bm K}$ and 
${ \bm p}'_{0 \bm K} = - p'_0 \hat{ \bm K}$. 

Inserting the amplitudes (\ref{qa-30}) and (\ref{qa-31}) 
into Eq. (\ref{qa-28}) and making use of Eq. (\ref{qa-29}),  
after some lengthy but elementary calculation we obtain 
\begin{eqnarray} 
\frac{ d \sigma_{fr}  }{ d^3 {\bm K}  
d^3 {\bm p}_1 d^3 {\bm p}_2 }  
&=& \Theta_{\bm K}    
\bigg[ 
\frac{ d \sigma^B_{\text{e-2e}}({\bm p}_{0 \bm K} ) }{ d^3 {\bm p}_1 d^3 {\bm p}_2 } \,   
\frac{ d \sigma^A_{ph}}{ d \Omega_{\hat{\bm K}}}  
\nonumber \\ 
&& \, \, \, \, \, \, \, \, \, 
 + \, \frac{ d \sigma^A_{\text{e-2e}}({\bm p}'_{0 \bm K} ) }{ d^3 {\bm p}_1 d^3 {\bm p}_2 } \, 
\frac{ d \sigma^B_{ph}}{ d \Omega_{\hat{\bm K}'}  } \bigg].   
\label{qa-32}  
\end{eqnarray} 
Here, $\frac{ d \sigma^B_{\text{e-2e}}({\bm p}_{0 \bm K}) }{ d^3 {\bm p}_1 d^3 {\bm p}_2 }$ and  
$ \frac{ d \sigma^A_{\text{e-2e}}({\bm p}'_{0 \bm K} ) }{ d^3 {\bm p}_1 d^3 {\bm p}_2 } $ are the electron impact ionization cross sections of atom B and A, respectively, 
which contain the energy conserving delta-function 
$\delta\big( \! \frac{p_1^2}{2 m_e} + \frac{p_2^2}{2 m_e}  + 
\frac{{\bm K}^2}{2 \mu } - \varepsilon_i - \epsilon_i - \hbar \omega \! \big)$. 
Further,  
$\frac{ d \sigma^A_{ph}}{ d \Omega_{\hat{\bm K}}}$ and 
$ \frac{ d \sigma^B_{ph}}{ d \Omega_{\hat{\bm K}'} }$ 
are the photo ionization cross sections of atoms A 
and B, respectively, differential in the emission angle (note that $\hat{\bm K}' = - \hat{\bm K}$ ), and 
\begin{eqnarray} 
\Theta_{\bm K} = \frac{ V_r }{ (2 \pi \hbar)^{3} } \, \, 
\bigg| \! \int \!\!d^3 {\bm R} \, 
\psi^*_{\bm K}({\bm R}) 
\, \frac{ 1 }{ R } \,\,  \psi_i({\bm R}) \bigg|^2.     
\label{qa-33}  
\end{eqnarray} 

The integration of the cross section (\ref{qa-32}) 
over the momenta ${\bm p}_1$ and ${\bm p}_2$ 
of the emitted electrons results in 
\begin{eqnarray} 
\frac{ d \sigma_{fr}  }{ d^3 {\bm K} }  
&=& \Theta_{\bm K} \,   
\bigg[ \sigma^B_{\text{e-2e}}( p_0 ) \, 
\frac{ d \sigma^A_{ph}}{ d \Omega_{\hat{\bm K}}}  
+ \sigma^A_{\text{e-2e}}( p'_0 ) \,
\frac{ d \sigma^B_{ph}}{ d \Omega_{\hat{\bm K}'}  } \bigg].   
\label{qa-34}  
\end{eqnarray} 
Here, $\sigma^A_{\text{e-2e}}(p'_0 )$ 
\big($ \sigma^B_{\text{e-2e}}(p_0) $\big) 
is the total cross section 
for ionization of atom A (B) by the impact of an electron incident with absolute momentum 
$ p'_0 $ ($p_0$) \cite{reduced-phase-volume}. 

\subsection{ The quantum approach and the reflection approximation }

The cross section (\ref{ga-10}) was derived using a very simple approach whereas the cross section (\ref{qa-32}) is an approximation 
to the cross section (\ref{qa-28}), obtained using the basic quantum consideration. However, the only difference between them is that the term 
$\frac{ Q_1 Q_2  }{ \mu \, K \, E^2_{K} } \, 
\, \bigg| \psi_i\bigg( \frac{ Q_1Q_2}{E_K} \hat{\bm K}  \bigg) \bigg|^2$ in Eq. (\ref{ga-10}) is replaced in 
Eq. (\ref{qa-32}) by the quantity $\Theta_{\bm K}$. 
Our test calculations show that, for a very broad range of energies $E_K$, these two quantities essentially coincide. 
In order to prove that this is not just a coincidence, 
let us consider the quantity $\Theta_{\bm K}$ 
in detail. 

The ground state of a very weakly bound 
A-B dimer is an $s$-state and can be written as 
\begin{eqnarray} 
\psi_i({\bm R}) = \frac{ \phi_0(R) }{ R } \, 
Y_{00}(\hat{\bm R}),   
\label{qa-35} 
\end{eqnarray} 
where the function $\phi_0(R)$ behaves at large $R$ as $\phi_0(R) \sim \exp( - \kappa_0 R)$ with 
$ \hbar^2 \kappa_0^2/2 \mu $ being the binding energy of the dimer. 

The state $\psi_{\bm K}({\bm R})$ 
of the A$^+$ - B$^+$ system can be taken, 
to an excellent approximation,  
as a Coulomb wave describing the relative  motion of the two singly charged fragments, which reads  
\begin{eqnarray} 
\psi_{\bm K}({\bm R}) & = & 
\frac{ 4 \pi }{ \sqrt{ V_r } } 
\sum_{l = 0}^\infty \sum_{m = - l}^{+l} i^l \, 
\exp(- i \delta_l) \, G_l(\kappa R) 
\nonumber \\ 
&& \, \, \, \times Y^*_{lm}(\hat{\bm K}) \, Y_{lm}(\hat{\bm R}),   
\label{qa-36} 
\end{eqnarray}
where 
\begin{eqnarray} 
G_l(\kappa R) & = & A_l  
\, (2 \kappa R)^l \, \exp(- i \kappa R) \,   
\nonumber \\ 
&& \, \times    _1F_1(l + 1 - i \eta, 2l+2, 2 i \kappa R).    
\label{qa-37} 
\end{eqnarray}
Here $\kappa = K/\hbar $, $\delta_l$ is the scattering phase, 
$ A_l = \frac{ \exp(- \pi \eta/2) \, |\Gamma(l+1+i\eta)| }{(2l +1)!}$ with $\Gamma(z)$ being the Gamma function, and  
$_1F_1(a,b,z)$ is the confluent hypergeometric function 
\cite{a-s}. Note that the function 
$G_l(\kappa R)$ is real and that the state  
$\psi_{\bm K}$ is normalized according to 
$ \langle \psi_{{\bm K}'} | \psi_{{\bm K}} \rangle = \frac{ (2 \pi \hbar )^3 }{ V_r } \, 
\delta^{(3)}({\bm K} - {\bm K}') $.  

Using Eqs. (\ref{qa-35})-(\ref{qa-36}) we obtain 
\begin{eqnarray} 
\int \!\!d^3 {\bm R} \, \, 
\psi^*_{\bm K}({\bm R})  
\, \frac{ 1 }{ R } \, \psi_i({\bm R}) 
\!\! & = \!\! &\frac{ 4 \pi }{ \sqrt{ V_e } } 
\, \exp(i \delta_0)   
\, Y_{00}(\hat{\bm K}) 
\nonumber \\ 
&& \!\!\! \times \!\!\! \int_0^{\infty} \! \! \! dR \,    
G_0(\kappa R) \, \phi_0(R).       
\label{qa-38}  
\end{eqnarray}  

\begin{figure}[h!]
\centering 
\hspace*{0.85cm} 
\vspace{-0.4cm}
\subfigure{\includegraphics[width=7.05cm,height=5.cm]{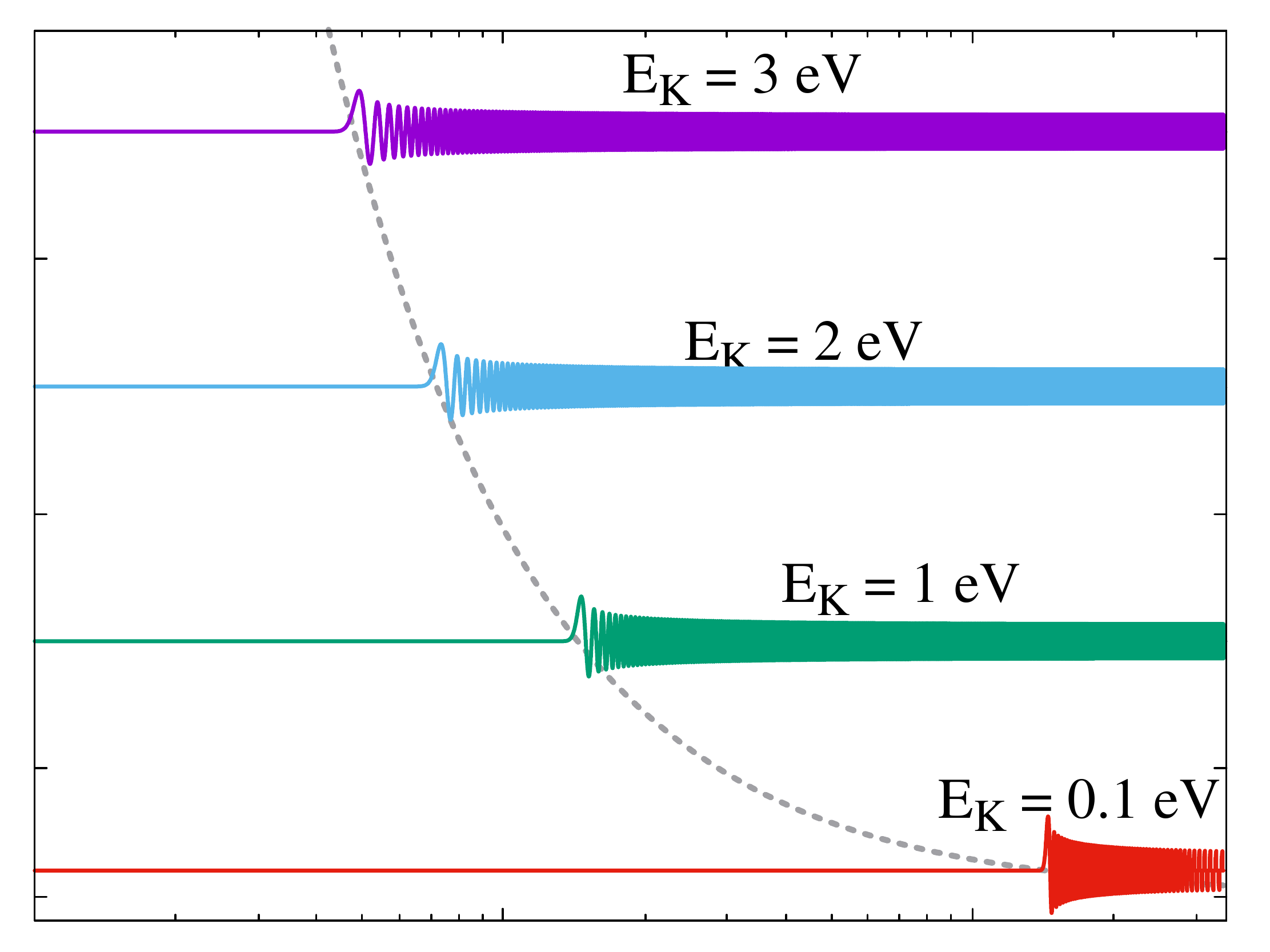}} 
\vspace{-0.4cm}
\subfigure{\includegraphics[width=8.5cm]{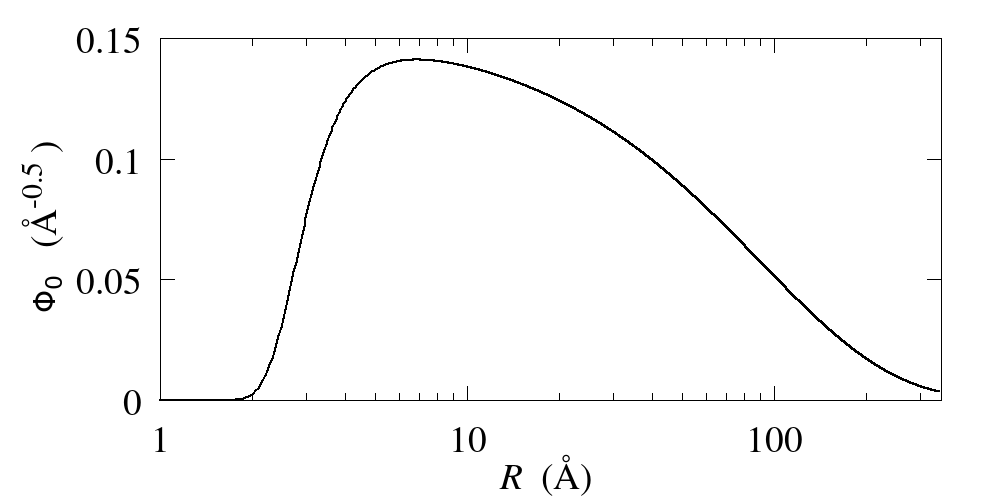}}
\caption{(a) The quantity $R \, G_0(\kappa R)$ 
for the $^4$He$^+$-- $^4$He$^+$ system, plotted as a function of the internuclear distance $R$ for different values of the kinetic energy release $E_K$. 
(b) The $R$-dependence of the function $\phi_0(R)$ 
for the ground state of the $^4$He$_2$. 
($\phi_0(R)$ was 
obtained by solving the Scr\"odinger equation with the interaction potential taken from \cite{potent_for_He2}.)  
For more explanations see text. }
\label{figure1} 
\end{figure}

In figure \ref{figure1} we illustrate the behaviour of the function $R \, G_0(\kappa R) $ for a few values of $E_K$ and also compare it with $ \phi_0(R)$. It follows from this figure that while the bound state function $ \phi_0(R)$ is very smooth, the continuum function $G_0(\kappa R) $ is extremelly highly oscillating, except when $R$ closely approaches the classical turning point $R_c = Q_1 Q_2/E_K$ and at $R < R_c$ where it exponentially decreases. Therefore, the main contribution to the integral in Eq. (\ref{qa-38}) 
is given by a small vicinity of $R_c$ (with the dimension of a few wave lengths $ 1/\kappa \ll 1$ a.u.) where the function $ \phi_0(R)$ is essentially a constant and one has 
\begin{eqnarray} 
\int_0^{\infty} \!\!\! \! \! \! \! dR \,    
G_0(\kappa R)  \phi_0(R) \!\! 
 & \approx & \!\! \phi_0(R_c) \! 
\lim_{\lambda \to + 0} \!\! 
\int_0^{\infty} \!\!\! \! \! \! \! dR  \, 
G_0(\kappa R) \exp(-\lambda R )  
\nonumber \\  
&=& \!\!\! A_0 \, \phi_0(R_c) \, \lim_{\lambda \to + 0} 
\frac{ 1 }{ \lambda + i \kappa} 
\nonumber \\ 
&& \!\!\!\!\!\times  _2F_1(1-i\eta, 1; 2; 2i \kappa/(\lambda + i \kappa)),  
\label{qa-39}  
\end{eqnarray}      
where $_2F_1(a, b; c; z)$ is the hypergeometric function 
\cite{a-s}. 
With the help of some properties of the hypergeometric function 
\cite{a-s}, \cite{hypgeom_func} one can show that  
\begin{eqnarray} 
&& \lim_{\lambda \to + 0} 
\frac{ 1 }{ \lambda + i \kappa} \, _2F_1(1-i\eta, 1; 2; 2i \kappa/(\lambda + i \kappa)) 
\nonumber \\ 
&& = \frac{ 1 }{ 2 \kappa \eta} 
\big(\exp(\pi \eta) - 1 \big).   
\label{qa-40}  
\end{eqnarray}      

Using Eqs. (\ref{qa-38}) - (\ref{qa-40}) and taking into account that $ A^2_0 = \exp(- \pi \eta) \, |\Gamma(1+i\eta)|^2 = 
\frac{ 2 \pi \eta }{ \exp(2 \pi \eta ) - 1 } $     
we finally obtain 
\begin{eqnarray} 
\Theta_{\bm K} =  \frac{ Q_1 Q_2 }{ \mu K E_K^2} \, 
\big | \psi_i(R_c \, \hat{\bm K}) \big |^2 \, \, 
\tanh(\pi \eta/2).  
\label{qa-41} 
\end{eqnarray}  
At $\eta \gg 1$ one has $\tanh(\pi \eta/2) 
\approx 1$ and the quantity $\Theta_{\bm K}$ coincides with the term  
$\frac{ Q_1 Q_2  }{ \mu \, K \, E^2_{K} } \, 
\, \bigg| \psi_i\bigg( \frac{ Q_1Q_2}{E_K} \hat{\bm K} \bigg) \bigg|^2$ in Eq. (\ref{ga-10}). 
Thus, the reflection approximation becomes very accurate when the Sommerfeld parameter $ \eta$ is large. 
Indeed, the uncertainty $\delta E_K$ in the "classical" value 
$E_K = Q_1 Q_2/R_c$ 
of the kinetic energy release 
 is caused by the width $ \delta R$ 
of the region of $R$, where the continuum wave function 
changes not very rapidly. This width is of the order of a few 
$ \kappa^{-1} $ and the relative uncertanty in the kinetic energy release $ \delta E_K/ E_K $ 
can be estimated as $ \delta E_K/E_K  = \delta R/R_c \simeq 
1/(\kappa R_c) \simeq 1/\eta$    
which means that at $\eta \gg 1$ 
the deviation of the kinetic energy release 
$E_K$ from its "classical" value 
$E_K = Q_1 Q_2/R_c$ is negligible. 

At this point a small clarification might be appropriate.  
On the one hand, since $\eta = \mu Q_1 Q_2/(\hbar K) $,  
the condition $\eta \gg 1$ would be better fulfilled 
for smaller values of the momentum $K$. 
On the other hand, this momentum 
still has to remain much larger than  
the recoil momenta $P_r = | {\bm p}_0 - 
M_A({\bm p}_1 + {\bm p}_1)/(M_A+M_B)| $ 
and $ P'_r = | {\bm p}'_{0} + 
\frac{ M_B }{M_A + M_B} ({\bm p}_1+{\bm p}_2) | $
which were neglected when obtaining 
the cross sections 
(\ref{qa-32}) and (\ref{qa-34}). 

\section{ Numerical results and discussion }

Numerical results reported in this section 
were obtained using Eq. (\ref{qa-34}) (the quantum approach) and 
Eq. (\ref{ga-11}) (the "geometric" approach).   

According to Eqs. (\ref{qa-34}) and (\ref{qa-41}) 
the angular distribution of 
(the relative motion of) the A$^+$ and B$^+$ 
fragments is determined by 
the photo cross sections 
$\frac{ d \sigma^A_{ph}}{ d \Omega_{\hat{\bm K}}}$ and 
$\frac{ d \sigma^B_{ph}}{ d \Omega_{\hat{\bm K'}}}$.  
Suppose that the fragmentation of the A-B dimer is triggered  by photo absorption from a beam of unpolarized photons which are incident along the $z$-axis. 
Assuming that both dimer atoms are initially in an 
$s$-state we have  
\begin{eqnarray} 
\frac{ d \sigma^A_{ph}}{ d \Omega_{\hat{\bm K}}} 
= \frac{ 3 }{ 8 \pi } \, \sigma^A_{ph} \, \sin^2(\theta_{\bm K}) 
\label{res_and_dis_1A} 
\end{eqnarray} 
and 
\begin{eqnarray} 
\frac{ d \sigma^B_{ph}}{ d \Omega_{\hat{\bm K'}}} 
= \frac{ 3 }{ 8 \pi } \, \sigma^B_{ph} \, \sin^2(\theta_{\bm K}), 
\label{res_and_dis_1B}  
\end{eqnarray}
where $\sigma^A_{ph}$ ($\sigma^B_{ph}$) is the total cross section for photo ionization of atom A (B)  and   
$\theta_{\bm K}$ is the polar angle of the relative 
momentum $\bm K$ of the heavy fragments.  
This means that the angular dependence of the 
fragmentation cross section 
$\frac{ d \sigma }{ d^3 \bm K} $ (given by Eq. (\ref{qa-34})) 
is simply proportional to $\sin^2 \theta_{\bm K}$.  

(Note that if the incident photons would be linearly polarized (say, along the $x$-axis) the angular dependence 
of the fragmentation cross section would be proportional 
to $\sin^2 \theta_{\bm K} \, \cos^2 \phi_{\bm K} $, 
where $ \phi_{\bm K}$ is the azimuthal angle of ${\bm K}$.)  

\vspace{0.25cm} 

\begin{figure}[h!]
\centering 
{\includegraphics[width=8.5cm,height=10.cm]{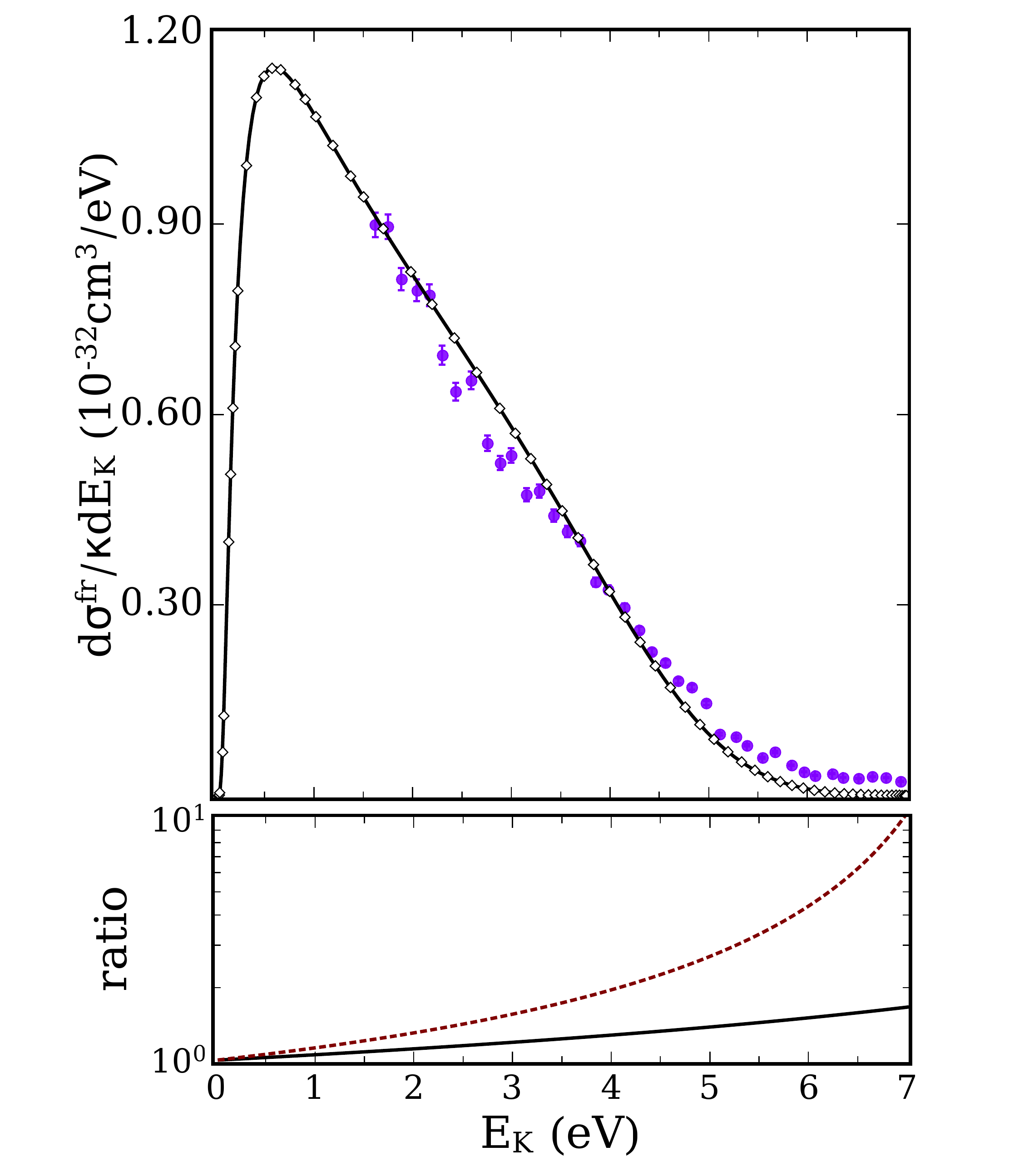}}
\put(-60,260){\color{black} {\fontsize{12}{8} 
\selectfont {\bf ({\it a})} }} 
\put(-60,73){\color{black} {\fontsize{12}{8} 
\selectfont {\bf ({\it b})} }} 
\caption{ ({\it a}) The cross section 
$ \frac{ d \sigma_{fr} }{ \kappa d E_K} $ ($\kappa = K/\hbar$) for the fragmentation of the $^4$He$_2$ dimer by absorption 
of a $63.86$ eV photon. 
Solid curve: results of our quantum calculations 
(using Eq. (\ref{qa-34})).  
Open diamonds: results of the "geometric" approach 
(Eq. (\ref{ga-11})).    
Solid circles with error bars: experimental data from \cite{He2_photon} normalized to our results. 
({\it b}) The ratio of the results 
obtained by i) fixing the  
value of $ \sigma^{He}_{\text{e-2e}} $ to  
$ \sigma^{He}_{\text{e-2e}}(\varepsilon_p = \hbar \omega - I_b)$, where $I_b \approx 24.59$ eV is the ionization potential of helium atom,     
and ii) taking into account the dependence of 
$ \sigma^{He}_{\text{e-2e}}$ on $E_K$. 
Solid curve: $\hbar \omega = 63.86$ eV. Dashed curve: 
$\hbar \omega = 57$ eV. For more explanations see text. }
\label{figure2} 
\end{figure} 

\subsection{ Fragmentation of He$_2$} 

We first consider the fragmentation of $^4$He$_2$
by absorption of a single photon with an energy of $63.86$ eV 
which was experimentally explored in \cite{He2_photon}. 

At this photon energy the photo electron momentum 
$ p_0 $ is $ \approx 1.7$ a.u. In our 
quantum approach we have assumed 
that $ p_0 R \gg \hbar $ which in this case 
is very well fulfilled starting with 
$ R \gtrsim 7$-$10$ a.u. 
($ R \gtrsim 4$-$5$ \AA).  

The value $ p_0 \approx 1.7$ a.u. 
also means that the recoil momenta $ P_r $ and $ P'_r $ are $ \simeq 1 $ - $2$ a.u. and, consequently, the corresponding recoil 
energy is $ \simeq 4$ - $16 $ meV. Therefore, when considering 
the kinetic energy release spectrum in the range 
$E_K \gtrsim 0.1$ eV the recoil effects can safely be neglected.    

In the derivation of the quantum transition amplitude we also assumed that the intermediate singly charged dimer remains "frozen" during the time interval $ T= R \, m_e/p_0 $, which the photo electron needs to propage from one dimer center to the other.   
The characteristic evolution time $ \tau $ 
of the singly charged dimer can be roughly estimated by using classical mechanics that yields 
$ \tau \approx \sqrt{ \frac{ \mu }{ \alpha } } \, \frac{ R^3 }{ Q } \, \frac{ \delta R }{ R }$, where $Q$ is the charge of the helium ion, 
$\alpha \approx 1.37$ a.u. is the static electric dipole polarizability of the ground-state helium atom and 
$\delta R $ is a (small) change in the size of the intermediate dimer. Assuming that $ \frac{ \delta R }{ R } = 10^{-2}$ 
(which means that the corresponding change in the kinetic energy release would be just about $1$ per cent) and taking into account that $\mu \approx 3600$ a.u. and 
$v \approx 1.7 $ a.u.       
we obtain that the ratio $ \tau/T \approx 
10^{-2} \sqrt{ \frac{ \mu }{ \alpha } } \, \frac{ R^2 \, v }{ Q } $ is much larger than $1$ already at $R \gtrsim 2$ \AA.   Thus, the intermediate $^4$He - $^4$He$^+$ dimer indeed does not have enough time to evolve. 

In figure \ref{figure2} we show results of 
our numerical calculations for the cross section 
$ \frac{ 1 }{ \kappa } 
\frac{ d \sigma_{fr} }{ d E_K } $ ($\kappa = K/\hbar $) for the fragmentation of $^4$He$_2$ by a $63.86$ eV photon 
in the range $  0.1$ eV $ \leq E_K \leq 7$ eV. 
In this range of $E_K$ the momentum 
$K$ varies between $ \approx 5$ and $ \approx 43$ a.u. 
being much larger than the recoil momenta 
for the most of the range,  
whereas the Sommerfeld parameter $\eta = Q_1 Q_2 \sqrt{ \mu/2 E_K }/\hbar  $ is always much larger than $1$.   
Therefore, the reflection approximation is valid 
and the quantum and "geometric" approaches 
(the cross sections (\ref{qa-34}) and (\ref{ga-11}), respectively) yield essentially 
the same results (see the figure).    

Our calculations were performed using     
the wave function for the ground state of 
the He$_2$ dimer which we    
obtained by solving numerically the Scr\"odinger equation with the interaction potential taken from \cite{potent_for_He2}.  

In figure \ref{figure2} we also compare our results with 
the experimental data from \cite{He2_photon}. Since the latter were not given on the absolute scale, we 
normalized them to the theory. This was done by setting  
the area under the experimental data 
(covering the interval $ 1.6 $ eV $ \leq E_K \leq 7$ eV) 
to be equal to the area under the theoretical curve 
(for the same energy interval).  
It is seen in the figure that our results 
are in reasonably good agreement with the experimental data.   

At a fixed photon energy, the dependence of 
the fragmentation cross section (\ref{qa-34}) 
on $E_K$ arises due to 
the quantity $\Theta_{\bm K}$, given by 
Eq. (\ref{qa-33}), and the cross section 
$ \sigma^{He}_{\text{e-2e}}$ for 
electron impact ionization which depends 
on $E_K$ mainly due to the reduced phase space volume for 
the final electron states caused by the energy outflow into the motion of the heavy reaction fragments. 
At a photon energy of $ \hbar \omega \approx 64$ eV  
the shape of the fragmentation 
cross section for  $ E_K \lesssim 7$ eV  
is determined mainly by $\Theta_{\bm K}$.  However, 
$ \sigma^{He}_{\text{e-2e}}$ also plays the important role  
accelerating the decrease of the fragmentation cross section in the interval $ 1 $ eV $\leq E_K \leq 7$ eV by 
a factor of $ \approx 1.6 $ 
(see figure \ref{figure2}b) and bringing 
the shape of the calculated spectrum in a noticeably better agreement with the experimental data. Moreover, it is worth noting that the effect 
of $ \sigma^{He}_{\text{e-2e}}$ on the shape would be even much stronger,  
had the photon energy be smaller (for instance, at $ \hbar \omega \approx 57$ eV it would accelerate the cross section decrease in the same interval of $E_K$ already by an order of magnitude, see figure \ref{figure2}b).    

\vspace{0.15cm} 
 
\begin{figure}[h!]
\centering 
{\includegraphics[width=8.cm,height=7.cm]{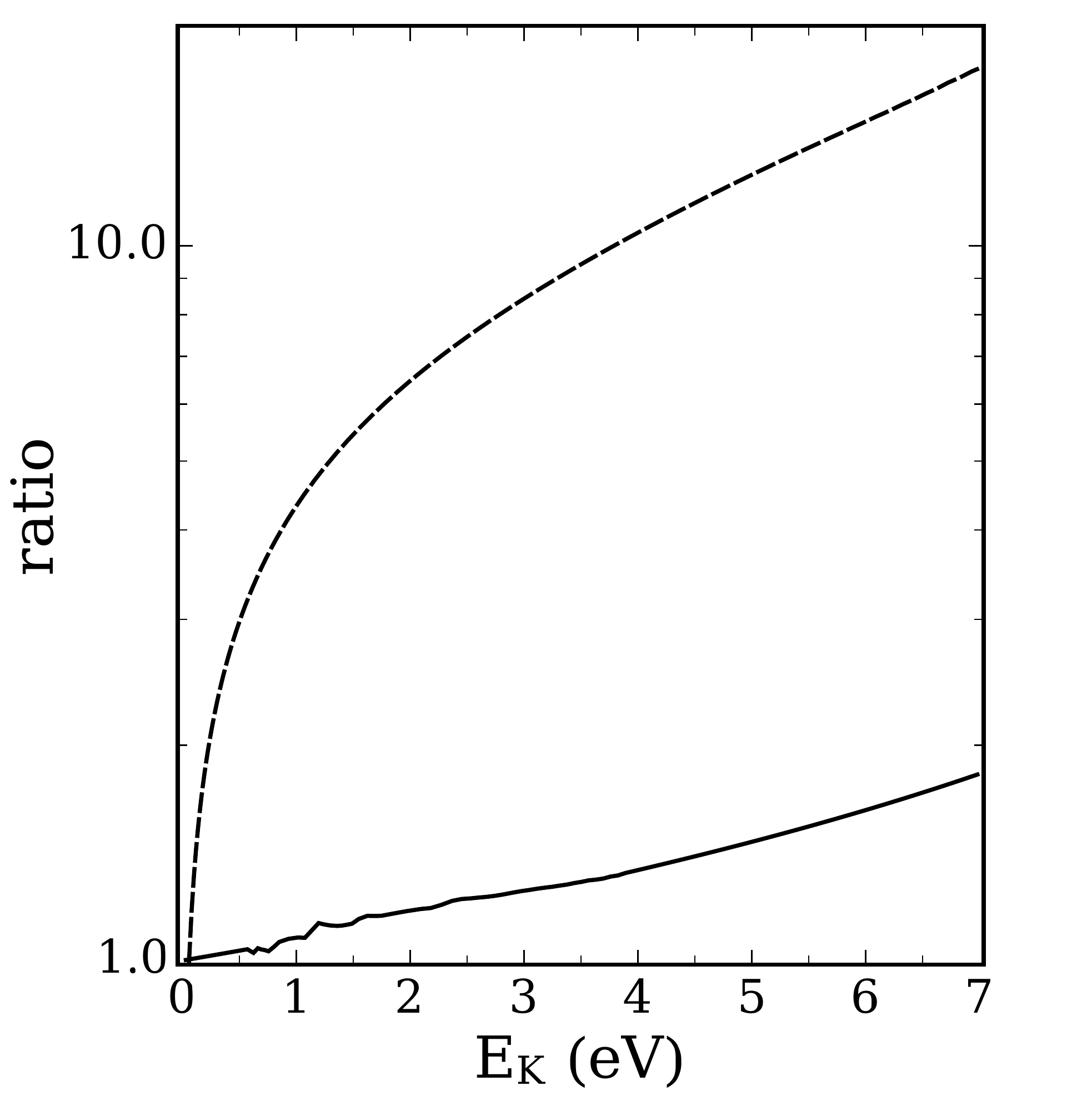}}
\caption{ Solid (dashed) curve: the ratio between 
the "quantum" ("classical") fragmentation probability,   
computed in \cite{He2_photon}, and our cross section  
$ \frac{ d \sigma_{fr} }{ \kappa d E_K} $. 
Both ratios were normalized to $1$ at $E_K \to 0$. For more explanations see text. } 
\label{figure-ratio} 
\end{figure} 
  
In addition to experimental results,  
two calculations for the shape of the kinetic energy release spectrum were made in \cite{He2_photon}.  
In one of them ("quantum") the probability 
$ P = \big | \int_{0}^{\infty} \!\!\!dR \Psi_i(R) f(R) \Psi_f(R) \big |^2$ was computed, where $\Psi_i(R)$ is the bound state of the He$_2$ dimer,  
$\Psi_f(R) $ is the state of the residual He$^+$ - He$^+$ system 
and the term $f(R) = \frac{ 1 }{ R }$ 
was {\it ad hoc} introduced; no derivation was given, which would show that $P$ does describe the spectrum shape. 
The other ("classical") 
employed the reflection approximation.

Concerning the correspondence between our results and the calculations of \cite{He2_photon} we note the following. 
The "quantum" probability $P$, computed in \cite{He2_photon},   
is proportional to the factor $\Theta_{\bm K}$ given by 
Eq. (\ref{qa-33}). However, 
$\Theta_{\bm K}$ is just 
one of the three terms entering the fragmentation cross section (\ref{qa-34})
and, if taken alone, 
determines neither the 
absolute values of the fragmentation cross section 
nor even the shape of its dependence on $E_K$ 
(see solid curve in figure \ref{figure-ratio}). 
Moreover, the shape of the energy distribution, predicted by the "classical" calculation  
of \cite{He2_photon}, and that, which follows from our consideration, 
differ even much stronger (see dashed curve in figure \ref{figure-ratio}). 

The reason for the latter is that the "quantum" and "classical" calculations of \cite{He2_photon} predict 
quite different spectrum shapes that led   
the authors of \cite{He2_photon} to a conclusion about 
the failure of the reflection approximation,  
attributing this to the very delocalized character 
of the He$_2$ ground state. Had this been true, 
it would have important consequences 
\cite{dimer_binding}.

However, our results do not support such a conclusion.  
In fact, as we have shown both numerically and analytically (see subsection II.C and also figure \ref{figure2}a),  
due to the quasiclassical character  
of the motion of the residual A$^+$ - B$^+$
system the "classical" reflection approximation    
works perfectly well 
for very large dimers including He$_2$ 
(as long as the recoil effects can be neglected). 

\vspace{0.15cm} 

In \cite{He2_photon} the angular distribution of the ionic fragments was measured. It was concluded there that it is proportional to $ \cos^2 \vartheta_{\bm K} $, where the angle $\vartheta_{\bm K} $ was counted from the polarization axis of the incident photon. This dependence agrees very well with 
the angular dependence of the 
fragmentation cross section 
$\frac{ d \sigma }{ d^3 \bm K} \sim 
\sin^2 \theta_{\bm K} \cos^2 \phi_{\bm K} $  
(where the angle $\theta_{\bm K}$ is counted from the photon propagation direction),   
which follows from our consideration 
(see the discussion at the beginning of Section III).  

\vspace{0.15cm} 

The authors of \cite{He2_photon} measured also the angular distribution of the emitted electrons 
and found that it agrees well with 
calculations  
for ionization of atomic helium by electron impact \cite{CCC} (assuming that the incident electron moves along the vector ${\bm K}$). In this respect we note that    
the fragmentation cross sections (\ref{ga-10}) and (\ref{qa-32})  
are a product of cross sections for 
atomic ionization by photo absorption and electron impact and the probability for a molecular transition. 
Besides, our consideration predicts that the photo electron emitted from one atom before colliding with the other atom moves along ${\bm K}$. All this clearly indicates that our theory describes the electron angular distributions as well.
     
\vspace{0.15cm} 

We conclude this part of the discussion by noting  that according to the calculation the total cross section for the fragmentation via the knockout mechanism is  
$ \approx 1.1 \times 10^{-22}$ cm$^2$.       

\subsection{ Fragmentation of Li-He dimers }

\begin{figure}[h!]
\centering 
{\includegraphics[width=8.5cm,height=8.cm]{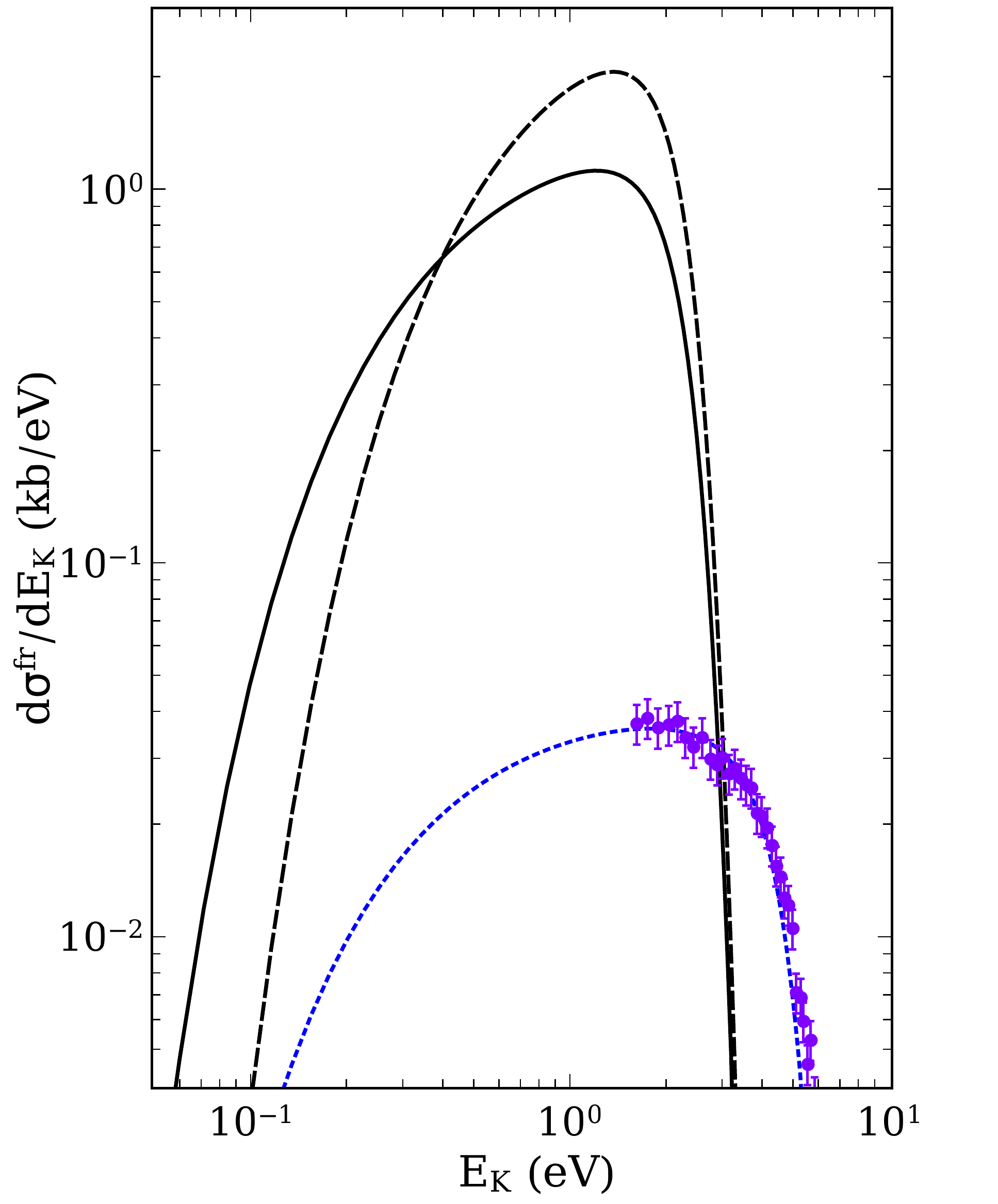}} 
\caption{ The cross section 
$ \frac{d \sigma_{fr} }{ dE_K } $ for the fragmentation 
of $^6$Li-$^4$He and $^6$Li-$^4$He dimers 
(solid and dashed curves, respectively) caused by 
absorption of a $45$ eV photon. For a comparison, 
results for the fragmentation of He$_2$  
by a $63.86$ eV photon are also shown. }
\label{figure3} 
\end{figure}

\begin{figure}[h!]
\centering 
{\includegraphics[width=8.5cm,height=5.cm]{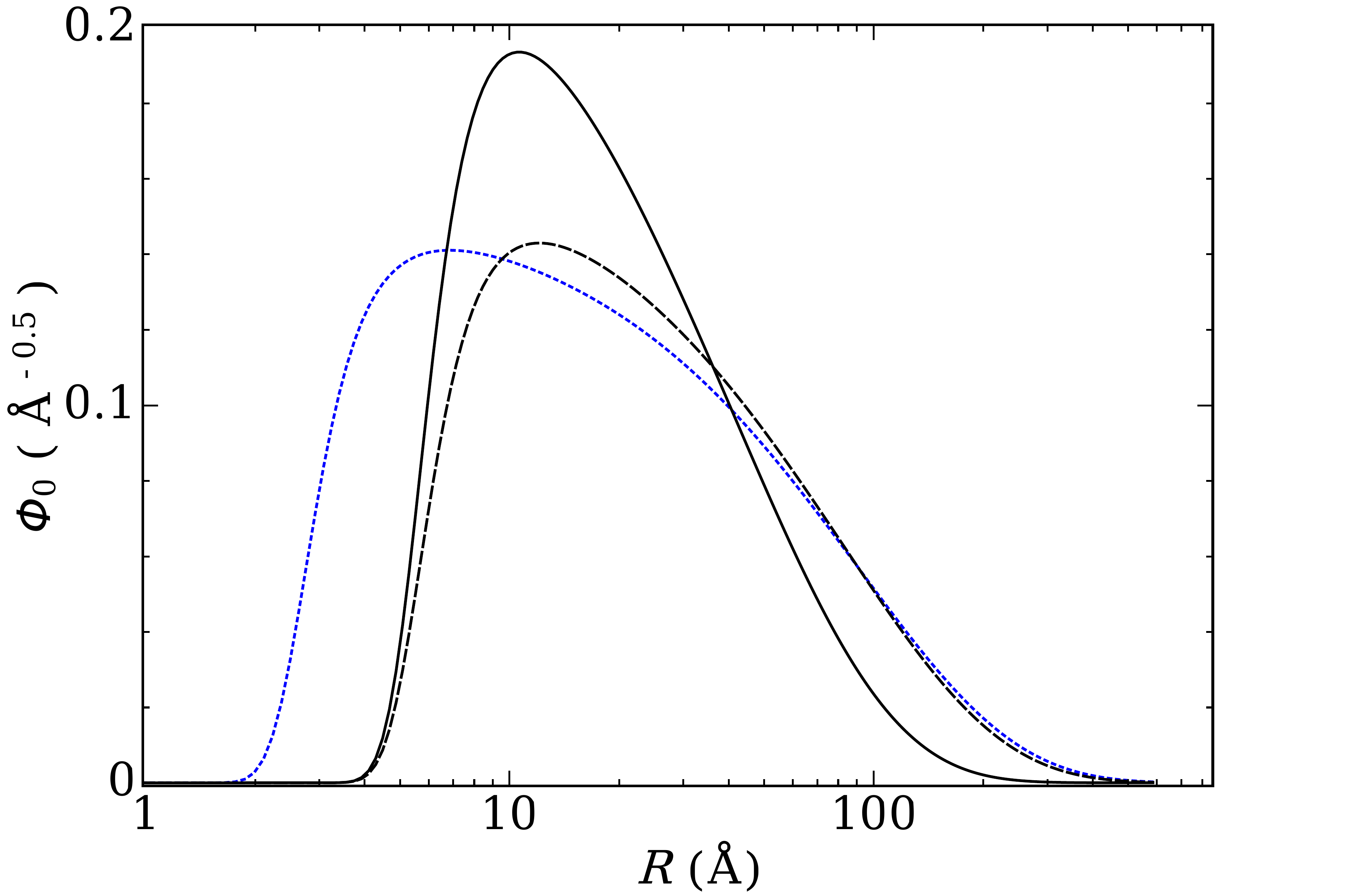}} 
\caption{ The function $\phi_0(R)$ for 
$^7$Li-$^4$He, $^6$Li-$^4$He 
and $^4$He$_2$ depicted by 
solid, dashed and dotted curves, respectively. }
\label{figure4} 
\end{figure}

\begin{figure}[h!]
\centering 
{\includegraphics[width=7.5cm,height=9.cm]{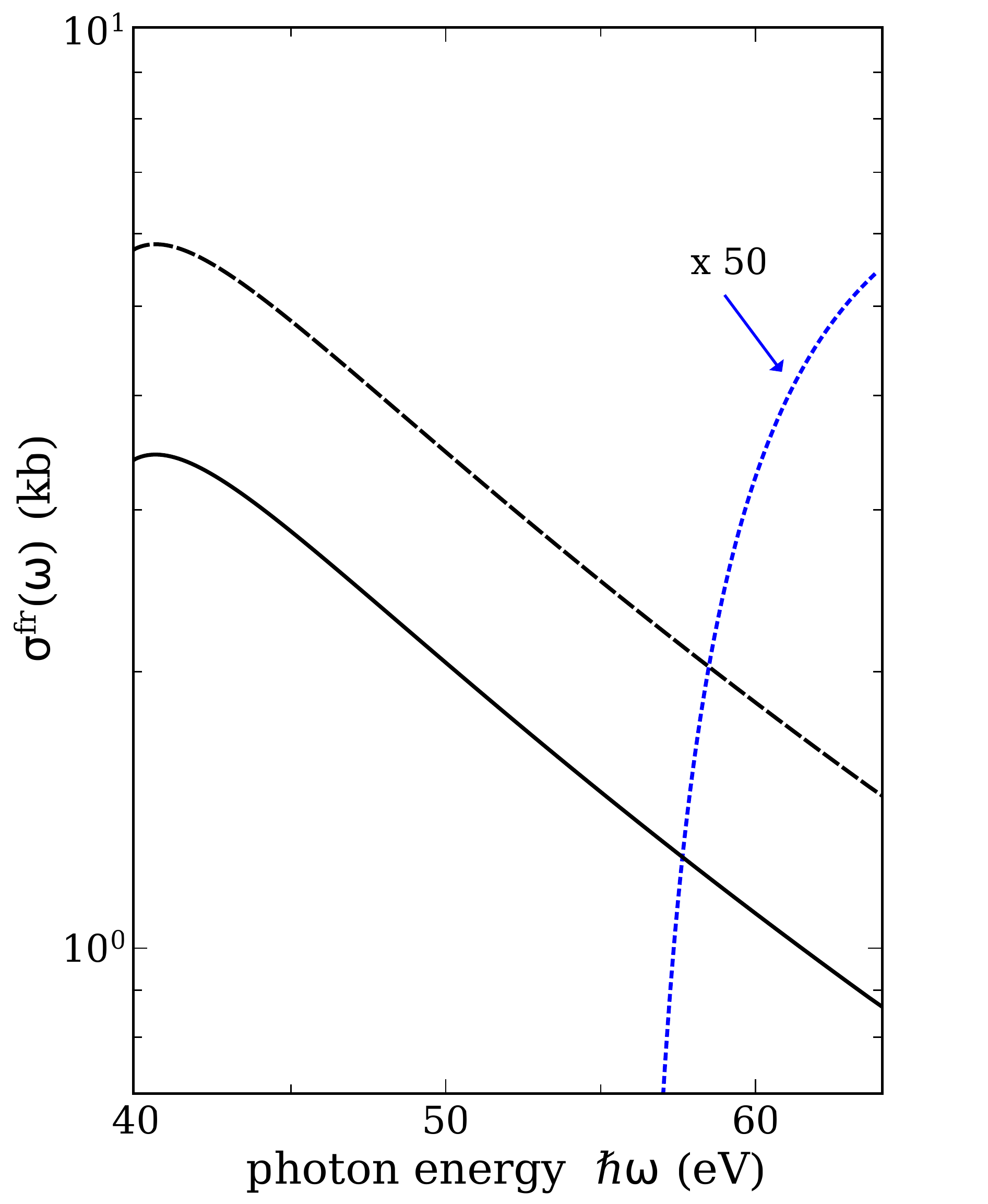}} 
\caption{ The calculated total fragmentation cross sections for $^7$Li - $^4$He (dashed curve), $^6$Li - $^4$He (solid curve)   
and $^4$He$_2$ (dotted curve) given as a function of the photon energy. The results for $^4$He$_2$ are multiplied by $50$. }
\label{figure5} 
\end{figure}

Now we turn to the fragmentation of the Li-He dimers. 
In order to breakup the dimers into Li$^+$ and He$^+$ ions 
the incident photon must have energy of at least $ |\varepsilon_i| + |\epsilon_i| \approx 30$ eV, where $ |\varepsilon_i| \approx 24.59$ eV and $ |\epsilon_i| \approx 5.39$ eV are the binding energy of atomic helium 
and lithium, respectively. 

Simultaneous ionization-excitation of He 
with the consequent ICD as well as 
photo emission from the $K$-shell of Li 
followed by Penning ionization and/or ETMD   
could also break the Li-He dimers into 
two singly charged ions. However, 
these processes have the energy threshold of 
$ \hbar \omega \approx 65.4$ eV 
and $ \hbar \omega \approx 64.4$ eV, respectively.  Therefore, for photon energies $ 30 $ eV $ \leq \hbar \omega < 64.4$ eV the knockout mechanism is the only one leading to the fragmentation into He$^+$ and Li$^+$ ions.    

As we have already seen, the photo fragmentation of an arbitrary (large) A-B dimer  
involves two reaction pathways -- 
photo ionization of A followed by electron impact ionization of B  and photo ionization of B followed by electron impact ionization of B. Unlike the photo fragmentation of the He$_2$ dimer, 
in the case of Li-He dimers the two reaction pathways possess quite different strength. Indeed, for photon energies in the range $ 30$ eV $ \leq \hbar \omega \leq 64$ eV 
the total photo ionization cross section of a He atom is 
by more than an order of magnitude larger than that of Li atom. Also, for the corresponding energies of the photo electron 
the total cross section for ionization by electron impact 
is much larger for Li. 
Therefore, in the case of Li-He dimers 
the fragmentation is strongly dominated by 
photo ionization of He and consequent photo electron impact ionization of Li.       

At photon energies $ 35$ eV $ \leq \hbar \omega \leq 64.4$ eV the momentum $ p_0 $ of the photo electron emitted from He varies between  $ 0. 87 $ and  $ 1.71 $ a.u.  meaning that at these energies the condition $p_0 R \gg \hbar $ is always very well fulfilled starting with  
$ R \gtrsim 8$ - $10$ \AA. 

Since the ground-state Li atom has quite a large static electric dipole polarizability 
($\alpha \approx 164$ a.u.) the intermediate 
Li - He$^+$ dimer evolves more rapidly than the He-He$^+$ system. 
In this case the ratio 
$ \tau/T \approx 
10^{-2} \sqrt{ \frac{ \mu }{ \alpha } } \, \frac{ R^2 \, v }{ Q } $ is much larger than $1$ at $R \gtrsim 7$ - $8$ \AA, 
where the intermediate Li - He$^+$ system  
can be viewed, to an excellent approximation, as "frozen".   

In figure \ref{figure3} we present our results for 
the cross section $ \frac{d \sigma_{fr} }{ d E_K} $ for 
the fragmentation of $^7$Li - $^4$He and 
$^6$Li - $^4$He dimers 
by absorption of a $45$ eV photon \cite{45-64}. 
In the range of $E_K$ shown in the figure 
the momentum $K$ is significantly larger than the momenta of the emitted electrons and, thus, the recoil effects in the kinetic energy release spectrum are weak. Besides, 
in this range the Sommerfeld parameter  
$\eta = Q_1 Q_2 \sqrt{ \mu/2 E_K }/\hbar $ is much larger than $1$. Therefore, the "geometric" approach employing the reflection approximation has a very good accuracy yielding practically 
the same results as the quantum treatment. 

In our calculation we used    
the wave functions for the ground state of 
$^7$Li - $^4$He and  $^6$Li - $^4$He dimers 
which we obtained by solving the Scr\"odinger equation with the interaction potentials taken from \cite{dimers-theor}.    

It follows from figure \ref{figure3} that both the intensity 
and the shape of the energy spectrum in the fragmentation of 
$^6$Li - $^4$He and $^7$Li - $^4$He differ markedly 
reflecting significant differences between 
the ground states of these dimers (see figure \ref{figure4}).  Besides, this figure also demonstrates dramatic differences between the cross sections for 
the Li - He dimers on the one hand and the $^4$He$_2$ dimer on the other. Now they are caused not only by the different shape of the dimer ground states (see figure \ref{figure4}) but also by 
quite different atomic cross sections 
for ionization by electron impact and photo absorption. 

In particular, while a much more rapid decrease in the energy spectrum at $E_K \gtrsim 2$ eV for $^6$Li - $^4$He compared to  $^4$He$_2$ (which both have about the same mean size) 
is caused by the fact that the ground state of  
$^6$Li - $^4$He is strongly suppressed already 
at $R \lesssim 5$-$6$ \AA (for the helium dimer 
this becomes the case only at $R \lesssim 2$ \AA), 
the difference in the overall intensity is related 
to the difference in the atomic cross sections.  
  
In figure \ref{figure5} we display results for the total fragmentation cross sections which clearly demonstrate that 
the photo fragmentation mechanism discovered in \cite{He2_photon} is much more efficient for Li-He dimers. As was already mentioned above, the reason is much larger atomic cross sections. Besides, the smaller size of $^7$Li - $^4$He leads to an additional enhancement of the fragmentation.

\subsection{Some additional remarks}

As was mentioned in the Introduction, 
at photon energies $ \hbar \omega > 65.4$ eV $ $ eV the fragmentation of the He$_2$ dimer into He$^+$ ions is on overall dominated by the ICD  
\cite{He2_photon_icd_exp}-\cite{He2_photon_icd_theor}. 
However, since the knockout and ICD mechanisms scale at $R \gg 1$ a.u.  as 
$1/R^2$ and $1/R^6$, respectively, the former could  nevertheless dominate fragmentation events with small kinetic energy release. Moreover, since the knockout mechanism is essentially instantaneous on the molecular time scale, it offers an opportunity to directly probe the structure of the ground state of the dimer.  
This could be especially the case for Li-He dimers where the knockout mechanism is much stronger than 
for He$_2$.  
   
\section{ Conclusions }

We have theoretically studied the the process of fragmentation of very large-size  dimers into singly charged ions by a single photon absorption, which was experimentally discovered for He$_2$ dimers in 
\cite{He2_photon}. In this process  
the fragmentation proceeds via photo ionization of either of the dimer atoms with the photoelectron moving towards the other atom and knocking out 
one of its electrons. The residual 
doubly charged system of atomic ions 
is unstable undergoing a Coulomb explosion.   

In order to describe this process we have developed two theoretical approaches which both make use of a large interatomic distances characteristic for very weakly bound dimers. One of them -- "geometric" -- is based on the consideration of the dimer's geometry,  
operates from the onset with atomic photo and electron impact ionization cross sections and employs the reflection approximation. 
The other -- "quantum" -- is a basic quantum approach, 
in which the process is considered by  
using the second order of the perturbation theory 
in the interaction of the "active" electrons of the dimer with the electromagnetic field and with each other. This approach enables one to obtain a very detailed information about the fragmentation process. 

We have also established the connection between these two approaches. In particular, they yield essentially identical results for the energy spectra of ionic fragments, provided i) there is little overlap between the final electron momentum spectra 
resulting from photon absorption by atom A and B, 
ii) the kinetic momentum of the relative motion of the ionic fragments is much larger than the recoil momenta 
and iii) the Sommerfeld parameter for 
the relative motion of the ionic fragments is much 
larger than $1$.     

We have applied the theory developed here 
to calculate the spectra for kinetic energy release and 
the total cross section for the fragmentation of  
$^4$He$_2$, $^7$Li-- $^4$He and $^6$Li-- $^4$He dimers.   
Our results for the helium dimer 
are in good agreement with available experimental data. 
For the Li-He dimers (where experimental data are still absent) our results predict that in this case  
the photo absorption is a much more efficient fragmentation mechanism than for He$_2$. The higher efficiency is first of all related to the fact that the cross section for the the electron impact ionization of the lithium atom are much larger than that for helium. 

The efficiency of the fragmentation mechanism studied here 
roughly scales with the dimer size as $1/R^2$ and thus has a much longer interaction range than the other fragmentation mechanisms (ETMD, Penning ionization and even ICD). Therefore, although at higher photon energies 
this mechanism is not the leading 
one in terms of the total fragmentation cross section (at least for the He$_2$ dimer) it can still dominate fragmentation events with small kinetic energy release.   
Besides, since on the molecular time scale this mechanism is essentially instantaneous, it can be used to directly probe the structure of the ground state of the dimer, including very large distances.   
This could be especially the case for Li-He dimers where 
it is much stronger than for He$_2$.  

\vspace{0.35cm} 

{\bf Acknowledgements}. B. Najjari gratefully acknowledges the support from the CAS President's Fellowship Initiative. 


\begin{thebibliography}{33}
\bibitem{i-a-a-1} 
In the inter-atomic Auger decay, considered in 
\cite{i-a-a}, a vacancy in an atom is filled by one of its electrons with the energy release transferred to a neighbour atom  that ionizes the latter.
\bibitem{He2+} The fragmentation 
of He$_2$ by fast protons via this mechanism was experimentally studied in \cite{He2_alpha-particle}.
\bibitem{1s2s2} A vacancy in the $K$-shell of Li can also be produced by photo excitation of a $K$-shell electron to states with higher principal quantum numbers that necessitates less photon energy 
(for instance, to form the 1s2s$^2$ state(s) one needs just about $51$ eV). However, in such a case, in order to form two singly charged ions,  
the transition of an electron from He to the $K$-shell of Li has to be accompanied by emission of two excited electrons from Li and thus involve the simultaneous interaction between three electrons that is expected to sharply reduce the ETMD efficiency.
\bibitem{en-cons} As before, in the energy conservation   
we have neglected the term  
$\frac{P_{cm}^2}{2 (M_A + M_B) }$ ($\simeq 1$ meV) related to the center-of-mass motion of the A$^+$--B$^+$ system  
and the tiny binding energy $E_i$ of the molecular state of the initial A-B dimer.
\bibitem{competition} This condition 
is fulfilled precisely for that range of photon energies,
where the process under consideration 
does not compete with other photo fragmentation processes.
\bibitem{reduced-phase-volume} Since a part of the photon 
energy goes into the kinetic energy of the residual ions,  
the phase space available for the emitted electrons is smaller than in the case of "normal" atomic e-2e process. Besides, the boundary conditions at infinity for the wave functions of the emitted electrons also change. These factors reduce the values of the electron impact ionization cross sections.
\bibitem{dimer_binding} For instance, 
it would completely invalidate results of the experiment 
\cite{dimer-binding-exp} where the reflection approximation was used for a precise determination of the binding energy 
of the He$_2$ dimer.
\bibitem{45-64} We note that the absorption of a $45$ eV photon 
by Li-He dimers leaves $\approx 15$ eV for the kinetic energies of the emitted electrons and the ionic fragments. This is about the same amount as when the He$_2$ dimer is fragmented by a 
$ 63.86 $ eV photon.
\end{thebibliography}
\end{document}